\documentclass[12pt]{article}
\usepackage{amsmath}
\usepackage{amssymb}
\usepackage{graphicx}
\usepackage[dvips]{color}
\usepackage{colordvi}

\catcode `@=11 \@addtoreset{equation}{section} \catcode `@=12
\begin{document}

\baselineskip=15pt

\def\gap#1{\vspace{#1 ex}}
\def\be{\begin{equation}}
\def\ee{\end{equation}}
\def\bal{\begin{array}{l}}
\def\ba#1{\begin{array}{#1}}  %% e.g. \ba{cc}
\def\ea{\end{array}}
\def\bea{\begin{eqnarray}}
\def\eea{\end{eqnarray}}
\def\beas{\begin{eqnarray*}}
\def\eeas{\end{eqnarray*}}
\def\del{\partial}
\def\eq#1{(\ref{#1})}
\def\fig#1{Fig \ref{#1}} 
\def\re#1{{\bf #1}}
\def\bull{$\bullet$}
\def\nn{\\\nonumber}
\def\ub{\underbar}
\def\nl{\hfill\break}
\def\ni{\noindent}
\def\bibi{\bibitem}
\def\ket{\rangle}
\def\bra{\langle}
\def\vev#1{\langle #1 \rangle} 
\def\lsim{\stackrel{<}{\sim}}
\def\gsim{\stackrel{>}{\sim}}
\def\mylsim{\lower .7ex\hbox{$\lsim$}}
\def\mygsim{\lower .7ex\hbox{$\gsim$}}
\def\mattwo#1#2#3#4{\left(
\begin{array}{cc}#1&#2\\#3&#4\end{array}\right)} 
\def\tgen#1{T^{#1}}
\def\half{\frac12}
\def\floor#1{{\lfloor #1 \rfloor}}
\def\ceil#1{{\lceil #1 \rceil}}

\def\mysec#1{\gap1\ni{\bf #1}\gap1}
\def\mycap#1{\begin{quote}{\footnotesize #1}\end{quote}}

\def\bit{\begin{item}}
\def\eit{\end{item}}
\def\benu{\begin{enumerate}}
\def\eenu{\end{enumerate}}
%%%%%%%%%%%%%%%%%%%%%%%%%%%%%%%%%%%%%%%%%%%%%%%%%%%%%%%%%%%%%%%%%%%
\def\a{\alpha}
\def\as{\asymp}
\def\ap{\approx}
\def\b{\beta}
\def\bp{\bar{\partial}}
\def\cA{{\cal{A}}}
\def\cD{{\cal{D}}}
\def\cL{{\cal{L}}}
\def\cP{{\cal{P}}}
\def\cR{{\cal{R}}}
\def\da{\dagger}
\def\de{\delta}
\def\tD{\tilde D}
\def\e{\eta}
\def\ep{\epsilon}
\def\eqv{\equiv}
\def\f{\frac}
\def\g{\gamma}
\def\G{\Gamma}
\def\h{\hat}
\def\hs{\hspace}
\def\i{\iota}
\def\k{\kappa}
\def\lf{\left}
\def\l{\lambda}
\def\la{\leftarrow}
\def\lar{\leftrightarrow}
\def\L{\Lambda}
\def\La{\Leftarrow}
\def\Lla{\Longleftarrow}
\def\Lra{\Longrightarrow}
\def\L{\Lambda}
\def\m{\mu}
\def\na{\nabla}
\def\nn{\nonumber\\}
\def\mm{&&\kern-18pt}  %\mm = my marker :)-
\def\om{\omega}
\def\O{\Omega}
\def\P{\Phi}
\def\pa{\partial}
\def\pr{\prime}
\def\r{\rho}
\def\ra{\rightarrow}
\def\Ra{\Rightarrow}
\def\ri{\right}
\def\s{\sigma}
\def\sq{\sqrt}
\def\S{\Sigma}
\def\si{\simeq}
\def\st{\star}
\def\t{\theta}
\def\tg{{\tilde g}}
\def\ta{\tau}
\def\ti{\tilde}
\def\tm{\times}
\def\tr{\textrm}
\def\Tr{{\rm Tr}}
\def\T{\Theta}
\def\up{\upsilon}
\def\Up{\Upsilon}
\def\v{\varepsilon}
\def\vh{\varpi}
\def\vk{\vec{k}}
\def\vp{\varphi}
\def\vr{\varrho}
\def\vs{\varsigma}
\def\vt{\vartheta}
\def\w{\wedge}
\def\z{\zeta}

\thispagestyle{empty}
\addtocounter{page}{-1}
\vskip-0.35cm
\begin{flushright}
TIFR/TH/09-14\\
%{\tt hep-th/******}
\end{flushright}
\vspace*{0.2cm} \centerline{\Large \bf Phases of one dimensional large
  $N$ gauge theory} 
\vspace*{0.2cm}
\centerline{\Large \bf in a $1/D$ expansion}
\vspace*{1.0cm} 
\centerline{\bf Gautam Mandal, Manavendra Mahato  and Takeshi Morita}
\vspace*{0.7cm}
\centerline{\it Department of Theoretical Physics,}
\vspace*{0.2cm}
\centerline{\it Tata Institute of Fundamental Research,} 
\vspace*{0.2cm}
\centerline{\it Mumbai 400 005, \rm INDIA}

\vspace*{0.5cm}
\centerline{\tt email: mandal, manav, takeshi@theory.tifr.res.in}

\gap2

\centerline{Draft date: \today}

\vspace*{0.8cm}
\centerline{\bf Abstract}
\vspace*{0.3cm}
\vspace*{0.5cm} 

We consider large $N$ Yang Mills theory with $D$ adjoint scalar fields
in $d$ dimensions for $d=0$ or 1.  We show the existence of a
non-trivial saddle point of the functional integral at large $D$ which
is characterized by a mass gap for the adjoint scalars.  We integrate
out the adjoint scalars in a $1/D$ expansion around the saddle
point. In case of one dimension which is regarded as a circle, this
procedure leads to an effective action for the Wilson line. We find an
analogue of the confinement/deconfinement transition which consists of
a second order phase transition from a uniform to a non-uniform
eigenvalue distribution of the Wilson line, closely followed by a
Gross-Witten-Wadia transition where a gap develops in the eigenvalue
distribution.  The phase transition can be regarded as a continuation
of a Gregory-Laflamme transition. Our methods involve large values of
the dimensionless 'tHooft coupling.  The analysis in this paper is
quantitatively supported by earlier numerical work for $D=9$.

\newpage
%%%%%%%%%%%%%%%%%%%%%%%%%%%%%%%%%%%%%%%%%%%%%%%%%%%%%%%%%%%%%%%%%%%
\tableofcontents

\section{Introduction and Summary}

Matrix models in low dimensions (especially 0 and 1) have served as
useful tools in many contexts. These include (i) $c\le 1$ matrix
models, which correspond to non-critical string theories
\cite{Brezin:1994eb}, (ii) large $N$ reduced models and their variants
\cite{Eguchi:1982nm,Krauth:1999qw}, (iii) BFSS matrix theory, which
corresponds to DLCQ of M-theory \cite{Banks:1996vh}, (iv) IKKT matrix
model of type IIB string theory \cite{Ishibashi:1996xs, Hotta:1998en},
(v) D0 brane black holes \cite{Itzhaki:1998dd, Kabat:2000zv,
  Kabat:2001ve, Anagnostopoulos:2007fw, Hanada:2008gy, Hanada:2008ez,
  Catterall:2008yz, Azeyanagi:2008mi}, (vi) KK reduction of
4-dimensional ${\cal N}=4$ SYM on $S^3$
\cite{Sundborg:1999ue,Aharony:2003sx,
  AlvarezGaume:2005fv,AlvarezGaume:2006jg}, (vii) The BMN matrix model
\cite{Berenstein:2002jq}, (viii) the matrix model of unstable D0
branes \cite{Klebanov:2003km, McGreevy:2003kb}, (ix) D branes on tori
\cite{Aharony:2004ig, Aharony:2005ew, Kawahara:2007fn, Azeyanagi:2009zf}, etc. In many
cases, one can regard these models as dimensional reduction of
large $N$ Yang-Mills theories. In models arising from D branes, the YM
theories are typically supersymmetric; however, in some situations the
theory is effectively described by the bosonic sector. The obvious
advantage of such a description, when it is possible, is that it is
easily amenable to numerical calculations, and in some fortunate
circumstances, to some powerful exact methods.

In this paper, we consider the following system for $d=0,1$:
\be
S=\frac1{g^2}  \int d^dx  \Tr \left( \frac{1}{2}  \sum_{I=1}^D
D_\mu Y^I D^\mu Y^I - \sum_{I,J} \frac14 [Y^I, Y^J]^2 \right),
\label{d-D}
\ee 
where $D_\mu$ refers to the covariant derivative $\del_\mu - i [A_\mu,
  .]$. $A_\mu, Y^I$ are $SU(N)$ matrices. For $d=0$ there is no gauge
field and the first term is absent. For $d=1$, there is a single gauge
field $A_0$ which is non-dynamical.

The action \eq{d-D} can be regarded as a dimensional reduction of
$D+d$ dimensional bosonic YM theory to $d$ dimensions. The specific
physical context we have in mind in the present paper is related to
case (ix) above, which is discussed in \cite{Aharony:2004ig,
  Aharony:2005ew, Kawahara:2007fn, Azeyanagi:2009zf} and reviewed below in Section
\ref{sec:review}.

In the $d=1$ case, we consider the dimension as a circle, of
circumference $\b$, and study the partition function and other
quantities as a function of $\b$. It was conjectured in
\cite{Aharony:2004ig, Aharony:2005ew, Kawahara:2007fn, Azeyanagi:2009zf}, on the basis
of numerical investigations, that the $d=1$ system exhibits a phase
transition which is analogous to the confinement/\-deconfinement
transition of ${\cal N}=4$ SYM on $S^3$.  The phase transition was
argued to be the weak coupling continuation of the black hole/black
string transition of the $d=2$ model. One of the motivations for the
present work was to understand the nature of the phase transition
analytically.

\gap1

Our results are briefly as follows: 

\begin{enumerate}

\item In the limit of large $D$, the model \eq{d-D} has a non-trivial
  saddle point characterized by a non-zero value of $\vev{\Tr Y^I
    Y^I}$. The large $D$ scaling is defined by keeping a modified
  'tHooft coupling $\tilde \l=\l D$ fixed.

\item In this limit, fluctuations around the above saddle point are
  suppressed by powers of 1/$D$. This enables us to develop a
  systematic expansion, in $1/D$, of the partition function and other
  quantities as a function of the radius of the circle.

\item In the $d=0$ case, the exact partition function is calculable up to
  $1/D$  {\em at finite $N$} (first shown in 
\cite{Hotta:1998en}), whereas in the $d=1$ case our results are
  obtained in the leading large $N$ limit.

\item Since the condensate provides a dynamically generated mass to
  the adjoint scalars $Y_I$, it is possible to explicitly integrate
  them out. In the $d=1$ case, this allows us to compute an effective
  action $S_{\rm eff}(W)$ for the Wilson line \be W = P \exp[i\int_0^\b dt
    A_0] \label{wilson} \ee in a $1/D$ expansion.

\item The $S_{\rm eff}$ computed in this fashion provides the first
  analytic evidence, in a $1/D$ expansion, for the phase transitions
  mentioned above.  It confirms the appearance of a double
  transition\footnote{We thank S.~Minwalla for a discussion on this
    point.}: (i) a second order phase transition characterized by the
  onset of non-uniformity in the eigenvalue distribution of $W$,
  followed by (ii) a third order Gross-Witten-Wadia (GWW) transition
  signalling the appearance of a gapped phase \cite{Gross:1980he,
    Wadia:1980cp, Wadia:1979vk}.  The appearance of a double
  transition is supported by the numerical works in
  \cite{Kawahara:2007fn, Azeyanagi:2009zf}\footnote{ We differ from
    \cite{Kawahara:2007fn, Azeyanagi:2009zf}, though, regarding the order of the two
    transitions. See Section \ref{sec 1/9} for details.}.  The phase
  transition temperatures $T_{c1}$ and $T_{c2}$, computed up to $1/D$,
  show excellent agreement with their results in the $D=9$ case, as
  shown in the following table (see Section \ref{sec 1/9} for more
  details and other comparisons):
%\begin{table}[h!]
\begin{center}
\begin{tabular}{l|c|c}
\hline & $ T_{ c1} $ & $ T_{ c2}$ \\ \hline 
Our result & 0.895 & 0.917 \\
Numerical result  & 0.8761 &
0.905 \\  \hline
\end{tabular}
\end{center}
% \end{table}

\item 
Our methods involve large values of the dimensionless 'tHooft
coupling.

\end{enumerate}

The large $D$ technique used in this paper for $d=0,1$ has earlier
been used in \cite{Hotta:1998en} in the $d=0$ context. 
 Our results for $d=0$ are in complete agreement with those of \cite{Hotta:1998en}, though
our method is slightly different (see Section \ref{sec 0d MM}
for details) in a way that enables a natural generalization 
to higher dimensions.  
A large $D$ expansion has also been used in certain lattice theories \cite{Drouffe:1983fv}.
Our methods also have some overlap with that of \cite{Kabat:2000zv,
  Kabat:2001ve} where a series of self-consistent equations (gap
equations) are introduced to determine various condensates and a GWW
type transition is suggested.

The paper is organized as follows. In Section \ref{sec:set-up} we set
up the main calculational method for the $d=1$ model. The method
consists of introducing an $SO(D)$-invariant dynamical field to get
rid of the commutator-squared interaction. This leads to an action
quadratic in the adjoint scalars with a dynamical mass term, which
allows us to integrate them out. We work out the example of the $d=0$
model in Section \ref{sec 0d MM} to test the method.  We show the
existence of a non-trivial saddle point in which a mass is generated
for the adjoints. We explain the nature of the large $D$ limit and
compute the partition function at finite $N$ in a $1/D$ expansion. In
Section \ref{sec d=1 contd} we come back to the $d=1$ model to derive
the effective action for the Wilson line in a $1/D$ expansion. We show
the existence of a second order phase transition followed by a GWW
transition. In Section \ref{sec:review} we provide a D brane
realization of our model, following earlier work \cite{Aharony:2004ig,
  Aharony:2005ew, Kawahara:2007fn, Azeyanagi:2009zf} which we briefly review. In Section
\ref{sec conclusion} we conclude with a discussion.

This paper arose as part of a larger project of exploring dynamical
black hole/black string transitions in terms of a dynamical large $N$
transition in a unitary matrix model \cite{progress}.

\section{\label{sec:set-up}The $d=1$ model: preliminaries}

We will set up the $d=1$ model in this section, test the formalism
with the simpler case of $d=0$ in Section \ref{sec 0d MM}, and
continue on to Section \ref{sec d=1 contd} to solve the $d=1$ model. A
D brane realization of the $d=0,1$ models is discussed in Section
\ref{sec:review}.

It is convenient to rescale the adjoint scalars $Y^I$ in \eq{d-D} to
$gY^I$, so that the action becomes 
\bea 
S \mm= \int_0^{\b} dt \, \Tr
\left( \sum_{I=1}^D \frac12 \left(D_0 Y^{I}\right)^2 - \sum_{I,J}
\frac {g^2}{4} [Y^I,Y^J][Y^I, Y^J] \right).
\label{matrixqm-action}
\eea We have assumed the theory to be on a circle, of circumference
$\b$, which can be interpreted either as a Euclidean time direction or
as a spatial circle. The covariant derivative is defined by $D_0 Y^I =
\del_t Y^I - i [A_0, Y^I]$.

The large $N$ limit is defined by keeping the 'tHooft coupling $\l =
g^2 N$ fixed, as $N\to \infty$. It is convenient to define the
following related dimensionless quantities
\begin{align}
\b_{\rm eff} &= \b\l^{1/3}= \b (g^2 N)^{1/3},
\label{beta-eff}
\\
\l_{\rm eff} &=\l \b^3 = \b_{\rm eff}^3.
\label{lam-eff}
\end{align}
For later convenience it is useful here to summarise  various
alternative definitions of coupling constant which we will use in this
paper in different contexts:
\gap2
\begin{center}
\begin{tabular}{l|l|l}
\hline
 $ \lambda $ & 'tHooft coupling & $g^2N$ \\
  $ \tilde{g}^2 $ &  'tHooft like coupling at large $D$ 
and finite $N$ & $g^2D$ \\  
  \lower .7ex\hbox{$\tilde{\lambda}$}  &  
'tHooft like coupling at large $D$ 
and large $N$ & $\lambda D$ \\
 $ \lambda_{\rm eff} $ &  dimensionless 'tHooft coupling 
&  $\lambda \beta^3$   \\ 
\hline
\end{tabular}
%\label{compare with num 1}
\end{center}

\gap2

\ni The last line refers only to the $d=1$ case.

We would like to explore the partition function 
\be
Z= \int {\cal D} A_0{\cal D}Y^I e^{-S},
\label{partition}
\ee
as a function of $\b_{\rm eff}$. We will also be interested
in the effective action $S_{\rm eff}(A_0)$, defined by
\be
\exp[- S_{\rm eff}(A_0)]= \int {\cal D}Y^I e^{-S}.
\label{s-eff}
\ee
It will turn out that $S_{\rm eff}$ depends only on the gauge-invariant
content of the gauge potential, namely on the eigenvalues of
the Wilson line $W$ \eq{wilson}.

%\subsection{The basic method} 

The first step in solving the model consists of making the action
(\ref{matrixqm-action}) quadratic in the $Y^I$ by introducing an
auxiliary field $B_{ab}$. Let us write $Y^I = \sum_{a=1}^{N^2-1}
Y_{a}^I \l_a$, where $\lambda_a$ are the generators of $SU(N)$. This
leads to an expression \be - \Tr[Y^I,Y^J][Y^I, Y^J]=
(Y^I_{a}Y^I_{b})M_{ab,cd}(Y_{c}^J Y_{d}^J),
\label{comm-sq}
\ee which is written in terms of $SO(D)$-invariant $Y$-bilinears,
where
\begin{align} 
M_{ab,cd} = -\frac{1}{4} \Bigl\{ \Tr[\l_a,
  \l_c][\l_b, \l_d] +(a\leftrightarrow b)+(c\leftrightarrow
d)+(a\leftrightarrow b,c\leftrightarrow d) \Bigr\}.
\label{def Mabcd}
\end{align}
Properties of the matrix $M$ are discussed in detail in Appendix
\ref{app Mabcd}.  Using the fact that $M$ is invertible\footnote{It
  might be puzzling, at first sight, that $M$ does not have a zero
  mode since \eq{comm-sq} clearly vanishes for special configurations
  of the `$Y^I$'s, e.g. when all `$Y^I$'s commute. The resolution is that
  $M$ has both positive and negative eigenvalues (see (\ref{eigen M}))
  and therefore has light-like vectors which, however, do not
  correspond to zero eigenvalues.  We thank Toby Wiseman for a useful
  discussion on this point.}
\footnote{The indefinite signature of $M$ (\ref{eigen M}) involves
  some subtlety in choosing the contour of the functional integral in
  \eq{gauss-trick}. E.g. we need to choose real contours for
  components of $B_{ab}$ along the positive eigenspace of $M$ and
  purely imaginary contours otherwise. The correct choice ensures
  finiteness of the normalization constant ${\cal N}$.\label{footnote N}}, we can write
\begin{align} 
Z= & {\cal N} 
\int {\cal D}B {\cal D}A_0 {\cal D}Y^I e^{-S(B,A_0,Y)},
\nn
S(B,A_0,Y) &= \int_0^\b dt\, \left[\frac12  \left(D_0 Y^{I}_a\right)^2 
- i\frac12  B_{ab} Y_{a}^I Y_{b}^I 
+ \frac{1}{4g^2} B_{ab} M^{-1}_{ab,cd}B_{cd} \right], 
\label{gauss-trick}
\end{align} 
with the following classical equation of motion for $B_{ab}$:
\be \frac1{g^2} M^{-1}_{ab,cd}B_{cd} = iY^I_{a}Y_{b}^I.
\label{b-y}
\ee In the above $1/{\cal N}\equiv \int {\cal D}B\, \exp[
-\int\ dt\,B_{ab}
  M^{-1}_{ab,cd}B_{cd}/(4g^2)]$, which we will ignore in the rest of
the paper since it involves only a numerical factor.  Since the action
\eq{gauss-trick} is quadratic in $Y^I$, we can formally integrate
them out, to get
\begin{align} 
Z= & \int {\cal D}B {\cal D}A_0 e^{-S_{\rm eff}(B,A_0)},
\nn
S_{\rm eff}(B,A_0) &= \int_0^\b\kern-5pt dt\,  
\frac{1}{4g^2} B_{ab} M^{-1}_{ab,cd}B_{cd} 
+ \frac{D}{2} \log {\rm det} \left((D_0^2)_{ab} + i B_{ab} \right).
\label{s-b-a0}
\end{align}  
For large $D$ \footnote{In much of this paper, we will treat $D$ as an
  arbitrarily specifiable parameter, except in the section dealing
  with comparison with D branes, where we put $D=9$. The precise
  scaling at large $D$ is defined in Eqns. \eq{tilde-g} and
  \eq{mod-tHooft}.  See also comments at the end of Section
  \ref{seff-leading}.}
  \footnote{ Although the large $N$ scaling is not apparent in
  \eq{s-b-a0}, the action as well as the measure admits a topological
  expansion in powers of $1/N$ for fixed $\l$, as is apparent from
  expressions such as (\ref{double-expansion}).} which scales as
$1/g^2$, we may expect the one-loop determinant to be comparable with
the classical term in \eq{s-b-a0} and hence modify the naive classical
solution $B_{ab}=0$. We will assume, and shortly justify, that
\eq{s-b-a0} admits a gauge-invariant time-independent solution, of the
form (see, e.g. Eqns. \eq{condensation 0d} and \eq{saddle-value-d=1})
\begin{align} 
\overline{B}_{ab} = i \triangle_0^2 \delta_{ab}.
\label{condensation assumption}
\end{align} 
The appearance of `$i$' is due to the fact, as we will see later, that
the solution corresponds to a saddle point in the complex plane. 
The condensate in terms of the original physical variables $Y^I$,
however, turns out to be real:
\begin{align} 
\langle \Tr Y^I Y^I \rangle
=\frac{N}{2g^2}\triangle_0^2,  
\label{condensation Y}
\end{align} 
where we have used \eq{b-y}, and also $M_{ab,cd}^{-1}
\delta_{cd}=\frac{1}{2N}\delta_{ab}$, which is derived in Eqn. \eq{m-inv-tr}.

To proceed, we write the $B_{ab}$ field as the sum of a constant trace
piece and the rest, as
\begin{align}
B_{ab}(t) = B_0 \delta_{ab} +g b_{ab}(t).
\label{fluct}
\end{align}
where $b_{ab}(t)$ satisfies $\int dt\, b_{aa}(t)=0$. We will show
below that $B_0$ has a saddle point value of the form $\overline{B}_0
= i \triangle_0^2$, consistent with \eq{condensation assumption}.
Substituting \eq{fluct} in the action in \eq{gauss-trick} we
get\footnote{Note the appearance of an effective mass term for the
  adjoint scalars $Y^I_a$ in the saddle point
  $\overline{B}_0=i\triangle_0^2$. In \cite{Aharony:2005ew} such a
  mass term is added by hand to integrate out the $Y^I$; here the
  mass term is dynamically generated.}
\begin{align} 
Z &= \int dB_0 \ {\cal D}A_0 {\cal D}b_{ab}\ {\cal D}Y^I e^{-S}, \quad
S = S_{0}+S_q+S_{int}, 
\end{align}
where\footnote{Strictly speaking, the last term in $S_q$ is a cubic
  term and should be regarded as an interaction. However, when we
  consider $B_0$ as an external parameter (unintegrated) we can regard
  this term as quadratic.},
\begin{align}
S_{0}&= \frac{\beta NB_0^2}{8g^2} , \nonumber \\
S_q &=\int_0^\beta dt\left( \frac{1}{4} b_{ab} M_{ab,cd}^{-1} b_{cd} 
+  \frac12 \left(D_0 Y_a^{I}\right) ^2
- \frac{i}{2} B_0 Y^{I2}_{a}\right) ,\nonumber \\
S_{int}&=- \int_0^\beta dt \left(\frac{ig}{2} 
b_{ab}Y^I_aY^I_b \right) .
\label{action 1dMM}
\end{align} 
Before proceeding to solve this model, let us discuss the $d=0$ matrix
model as a partial test of our formalism.

\section{The $d=0$ model}
\label{sec 0d MM}

The $d=0$ model is defined by a partition function\footnote{Most of
  the results in this section are in \cite{Hotta:1998en} who have
  discussed this model earlier in the context of the IKKT matrix
  model. Our method, however, is slightly different, especially in the
  way we distinguish between the diagonal and the off-diagonal
  fluctuations of the auxiliary field $B_{ab}$ (see, e.g. \eq{fluct}
  and \eq{0d partition function}) which allows for a natural
  generalization to the $d=1$ model.}
\begin{align} 
Z= \int dY^I \exp[- S], \quad S =  
 - \frac{g^2}4 \Tr \sum_{I,J}[ Y^I, Y^J]^2.
\label{d=0}
\end{align} 
Here $Y^I$ are $SU(N)$ hermitian matrices which are
normalized the same way as in \eq{matrixqm-action}. A D brane
interpretation of this model is discussed in Section \ref{sec:review}.

Like in the previous section for $d=1$, we can rewrite \eq{d=0} as
\begin{align} 
Z &= \int dB_0  db_{ab}dY^I \exp[-S],
\\ S &= S_0 +S_q+S_{int}, \nonumber 
\\ S_0&= \frac{NB_0^2}{8g^2}, ~~ S_q= \frac{1}{4}
b_{ab} {M}_{ab,cd}^{-1} b_{cd} -i \frac{B_0}{2}Y^{I2}_{a}, ~~
S_{int}= -\frac{ig}{2}b_{ab}Y^I_aY^I_b.
\label{d=0-big}
\end{align} 
Integrating out the $Y^I$ gives us the $d=0$ analogue of
\eq{s-b-a0} where $B_{ab}$ is split into $B_0$ and $b_{ab}$:
\begin{align}
Z&=\int dB_0 db_{ab} e^{-S_{\rm eff}(B_0, b_{ab})}, \nn S_{\rm eff}(B_0,b_{ab})
&= \frac{NB_0^2}{8g^2} + \frac{1}{4} b_{ab} {M}^{-1}_{ab,cd}b_{cd}
+ \frac{D}{2} \log {\rm det} \left(i B_0\delta_{ab} + 
i gb_{ab} \right).
\label{s-b-b0}
\end{align}  

\subsection{The large $D$ limit}\label{sec:large-D}

Let us make a formal Taylor expansion of $S_{\rm eff}$ in
powers of $b_{ab}$:
\begin{align}
S_{\rm eff}(B_0, b_{ab}) &= S^{(0)}(B_0) +\frac12 S^{(2)}_{ab;cd}(B_0)
b_{ab}b_{cd} + S_{\rm eff}^{int} ,
\nn S^{(0)}(B_0) &= \frac{NB_0^2}{8g^2}
+ \frac{D}{2} \log {\rm det} \left(i B_0\delta_{ab}\right) ,
\nn
S^{(2)}_{ab;cd}(B_0) & = \frac12 M^{-1}_{ab,cd} + \frac{g^2 D}{2}
\frac{\del^2 \log {\rm det}\left(i B_0\delta_{rs} + i \bar b_{rs}
  \right)}{\del \bar b_{ab}\del \bar b_{cd}}\Bigr|_{\bar b_{rs}=0}, 
\nn S_{\rm eff}^{int} &= O(b^3),
\label{large-D-logic}
\end{align}
where in defining $S^{(2)}$ we have used $\bar b_{ab}
\equiv g b_{ab}$ in order to exhibit the $g$ dependence explicitly.
There is no linear term in the above Taylor expansion  since
$b_{aa}=0$. Let us define a large $D$ limit by keeping
\begin{align}
{\tilde g}^2 \equiv g^2 D,
\label{tilde-g}
\end{align}
fixed. It is easy to see that
\begin{align}
S^{(0)}(B_0)= O(D),\quad S^{(2)}_{ab;cd}(B_0) = O(1),\quad
S_{\rm eff}^{int} = O(1/D).
\label{orders-D}
\end{align}
Let us now do the integral in \eq{s-b-b0} over $b_{ab}$, to give
\begin{align}
Z=& \int dB_0 \exp[-{\cal S}(B_0)],
\label{z-b0} \\
\frac{{\cal S}(B_0)}{D}=& {\cal S}_0(B_0,\tilde g) + 
\frac1{D}{\cal S}_1(\tilde g)
+ O\left( \frac1{D^2} \right) ,
\end{align}
where 
\begin{align}
{\cal S}_0(B_0,\tilde g) &= \frac1{D} S^{(0)}(B_0)
= \frac{NB_0^2}{8{\tilde g}^2} +
\frac{1}{2} \log {\rm det} \left(i B_0\delta_{ab}\right)
,\nn
{\cal S}_1(B_0, \tilde g) &= 
\frac12 \log {\rm det} S^{(2)}_{ab;cd}.
\label{cal-s0-1}
\end{align}
The first `log det' is essentially a 1-loop integral over $Y^I$, while
the second `log det' is a 1-loop integral over $b_{ab}$ (see Figure
\ref{fig D orders}(a) and (b)).  In Appendix \ref{sec:s-s} we will
present an explicit computation of these quantities. We find (see
\eq{finite N d=0})
\begin{align}
{\cal S}_0(B_0)&=
\frac{N B_0^2}{8{\tg}^2 } +\frac{(N^2-1)}{4}\log
 \left(- \frac{B_0^2}{\tg^2N } \right),
\nn
{\cal S}_1(B_0)&= \frac{N^2-1}{2}
 \log\left(1-\frac{\tg^2N}{B_0^2} \right) 
 +\frac{N^2(N+1)(N-3)}{8} \log\left(1-
\frac{2\tg^2}{B_0^2}
 \right) \nn
& +\frac{N^2(N-1)(N+3)}{8}
 \log\left(1+ \frac{2\tg^2}{B_0^2} \right).
\label{cal-s0-1-val}
\end{align}
In the $d=1$ case, an explicit finite $N$, large $D$ result such as
Eqn. \eq{cal-s0-1-val} is difficult to obtain, but
we will derive an analogue of Eqn. \eq{s-s} below. Furthermore, as
remarked at the end of Section \ref{seff-leading}, for $d=1$ we will
not take the strict $D=\infty$ limit since criticality involves $1/D$
effects.

\subsection{Large $D, N$ limit}

It is easy to see that both ${\cal S}_0$ and ${\cal S}_1$
admit a 'tHooft-like expansion in which 
\begin{align}
\tilde \l = \l D = g^2 ND= \tg^2N ,
\label{mod-tHooft}
\end{align}
is kept fixed.  We obtain an expansion of the sort
\begin{align}
\frac{{\cal S}}{DN^2}= \left( {\cal S}_{0,0} + \frac{1}{N^2} {\cal
  S}_{0,1} + \cdots \right) + \frac1{D}\left( {\cal S}_{1,0} +
\frac{1}{N^2} {\cal S}_{1,1} + \cdots \right) + \cdots.
\label{double-expansion}
\end{align}
In the diagrammatic evaluation described in the Appendices, such an
expansion indeed corresponds to a topological expansion.  Explicitly,
from \eq{cal-s0-1-val}, we get
\begin{align}
\frac{\cal S}{DN^2} = & \frac{B_0^2}{8\tilde \lambda} +\frac{1}{4}\log
  \frac{-B_0^2}{\tilde{\lambda}} + \frac1{D}
  \left\{\frac{\tilde{\lambda}}{B_0^2} - \frac{1}{2}\left(
  \frac{\tilde{\lambda}}{B_0^2}\right)^2
  +\frac{1}{2}\log\left(1-\frac{\tilde{\lambda}}{B_0^2}
  \right)\right\}   
\nn &  + O\left(1/D^2\right) +
  O\left(1/N^2\right).
\label{s-s}
\end{align}

\subsection{The saddle point}

We are left with evaluating \eq{z-b0}. Because of the appearance of an
overall factor of $N^2$ in \eq{s-s}, we can evaluate \eq{z-b0} using a
saddle point method (see a similar calculation presented in Appendix
\ref{saddle-point} in a toy example). The saddle point value is given
by 
\be \overline{B}_0 = i \triangle_0^2,  \quad \triangle_0^4 = 2\tilde\l
\left(1 + \frac7{3D}\right) + O(1/D^2).
\label{condensation 0d}
\ee The same result is also derived in \cite{Hotta:1998en} in a
slightly different manner. \cite{Hotta:1998en} also performed a
numerical analysis which agrees with the above analytical calculation
and also with the numerical calculations of \cite{Aharony:2004ig,
  Aharony:2005ew}.

Using the above saddle point, we get the free energy,
\begin{align} 
F=-\frac{\log Z}{DN^2}= -\frac{1}{4}+ \frac{\log 2}{4} +\frac1{D}
\left(- \frac{5}{8}+\frac{1}{2}\log \frac{3}{2} \right) + O\left(\frac
1 {D^2} \right) .
\end{align} 

\section{The $d=1$ model: result}
\label{sec d=1 contd}

After gaining some experience with the $d=0$ matrix model, we now
return to the more involved case, the $d=1$ model. We start with
\eq{action 1dMM}, and as with Eqns. \eq{z-b0} and \eq{0d partition
  function}, we first integrate out the $Y^I$ and the $b_{ab}$ to
obtain
\begin{align}
Z = \int dB_0 {\cal D}A_0 e^{- {\cal S}(B_0, A_0)}, 
\label{d=1-part}
\end{align}
where
\be
e^{- {\cal    S}(B_0, A_0)} = \int {\cal D}b_{ab} {\cal D}Y^I e^{-S},
\label{d=1-seff}
\ee with $S$ defined as in \eq{action 1dMM}.  Different from the
previous section, we consider large $N$ case only in this section.

It is convenient to parametrize $A_0$ by choosing a gauge in which
$A_0$ is time-independent and is also diagonal: $A_{0ij}=
\alpha_{i}\delta_{ij}$. The gauge-invariant content of $A_0$ is then
given by the moments
\begin{align}
u_n = \frac{1}{N} \Tr W^n= \frac{1}{N} \sum_{i=1}^N e^{in\beta \alpha_i} ,
\end{align} 
where $W$ is the Wilson loop operator, defined in \eq{wilson}.  The
above gauge fixing gives rise to a Faddeev Popov Jacobian (See
\cite{Aharony:2003sx})
\begin{align} 
{\cal D}A_0=\prod_i d\alpha_i e^{-S_{FP}},\quad S_{FP}=N^2\sum_n 
\frac1{n} |u_n|^2.
\label{FP}
\end{align} 

It is convenient to parametrize $B_0= i \triangle^2$, since the saddle
point value will be real in terms of $\triangle$ as in the $d=0$ case
(\ref{condensation 0d}).  From now on, we will denote ${\cal S}(B_0,
A_0)$ as ${\cal S}(\triangle, \{u_n\})$.

Note that there is a Jacobian involved in changing from the
integration measure over the eigenvalues $\a_i$ to the
integration measure over $u_n, \bar u_n$; however, it is
$O(N)$ and is hence subleading compared to the classical
action which is $O(N^2)$ \cite{Jevicki:1979mb}. Since in this
section we will be concerned with the leading term
in the $1/N$ expansion, we will ignore this Jacobian.

\subsection{Computation of ${\cal S}(\triangle, \{u_n\})$ in
leading large $D$}\label{seff-leading}

As in Section \ref{sec 0d MM}, we can ignore the interaction $S_{int}$
in the large $D$ limit. Hence the leading result of the effective
action is obtained by integrating out the $Y^I$ from $S_q$ in \eq{action
  1dMM}.

We can integrate out $Y^I$ by using the propagator studied in Appendix
\ref{app propagator} (following \cite{Aharony:2003sx}) and obtain
\begin{align}
\frac{D}{2} \log\left(\hbox{det} \left( - D_0^2 + \triangle^2 \right)
\right)=& \frac{DN^2 \beta \triangle}{2} -D\sum_{n=1}^\infty
\frac{x^n}{n}| u_n |^2.
\end{align}
Here $x=e^{-\beta \triangle}$ and we have ignored $1/N$ terms and
irrelevant constant terms.  We also ignored a temperature dependent
divergent term.

Combining the above equation with $S_0$ from \eq{action 1dMM}, and
adding the contribution from \eq{FP}, we get
\begin{align}
\frac{{\cal S}(\triangle,\{u_n\})}{D N^2}= 
-\frac{\beta \triangle^4}{8 \tilde \l} +
\frac{\beta \triangle }{2} +
\sum_{n=1}^\infty
\left(
\frac{1/D- x^n}{n} \right)  |u_n|^2.
\label{effective action d=1}
\end{align}
The $1/D$ term comes from \eq{FP}. The reason it is kept here is that
near the critical temperature this term will turn out to be more
significant than other $O(1/D)$ terms from $S_{int}$ which we will
encounter in the next subsection.

Our task is to evaluate \eq{d=1-part}, with ${\cal S}(B_0, A_0) ={\cal
  S}(\triangle,\{u_n\}) $ given above. It is useful to first perform
the integral over $B_0$, using a saddle point method similar to
Section \ref{sec 0d MM}. In other words, for fixed external $u_n$, let
us now solve the saddle point equation
\begin{align} 
-\frac{\triangle^3}{2\tilde{\lambda} }+\frac{1}{2} +\sum_{n=1}^\infty
e^{-n\beta \triangle} |u_n|^2=0.
\label{d=1 saddle point equation}
\end{align} 
It is difficult to solve this equation for $\triangle(\{u_n\})$
exactly.  However, for small $u_n$, it is possible to solve it in a
power series in the $u_n$. This leads to the following saddle point
solution
\begin{align}
\triangle_0(\{u_n\})= {\tilde \l}^{1/3}
\left(1 + \frac23 \sum_{n=1}^\infty {\bar x}^n|u_n|^2 \right) 
+\cdots,
\label{saddle-value-d=1}
\end{align}
where 
\begin{align}
\bar x= \exp[-\b {\tilde \l}^{1/3}].
\label{bar-x}
\end{align}
Substituting \eq{saddle-value-d=1} in \eq{effective action d=1}, we
get a Landau-Ginzburg type effective action for the $u_n$:
\begin{align}
\frac{{\cal S}(\{u_n\})}{DN^2} &= \frac{3}{8} \b {\tilde \l}^{1/3}
+ a_1 |u_1|^2 + b_1 |u_1|^4 + \sum_{n=2}^\infty a_n |u_n|^2+ \cdots,  
\nn a_n &=
\frac1{n}\left(1/D - {\bar x}^n\right), \nn b_1 &= \frac13 \b {\tilde
  \l}^{1/3}{\bar x}^2,
\label{LG}
\end{align}
where the $\cdots$ involve other $u^4_n$ terms for $n>1$, which are
ignored for reasons stated below.

Let us analyze the phase structure of the theory by using \eq{LG} (see
Figure \ref{fig phase}).\footnote{We will show below that inclusion of
  higher loop terms does not change the nature of phase transitions,
  although it changes the critical temperature and numerical values of
  various thermodynamical quantities.} Our analysis will be similar to
\cite{Aharony:2003sx}. Note that for $\bar x < 1/D$ all `$a_n$'s are
positive. This implies that $\{u_n=0$ $\forall$ $n=1,2,...\}$ is a
minimum of the potential.\footnote{The issue of whether it is a local
  or a global minimum is more subtle, and depends on details of higher
  order terms in \eq{LG}. We will argue below that in a $1/D$
  expansion the higher order terms can be ignored and $u_n=0$ is a
  global minimum.}

\begin{figure}
\begin{center}
\includegraphics[scale=0.75]{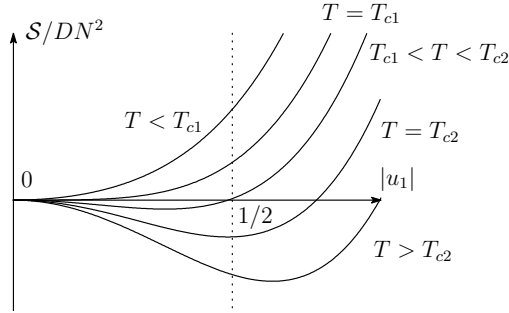}
\caption{Phase transitions: ${\cal S}$ vs $|u_1|$ (see \eq{LG},
  \eq{LG'}). As $T$ crosses $T_{c1}$, $u_1$ becomes tachyonic and
  there is a second order phase transition which signals an onset of
  non-uniformity in the eigenvalue distribution $\rho(\a)$.  At
  $T=T_{c2}$, characterized by a potential minimum at $|u_1|=1/2$, a
  gap develops in the eigenvalue distribution, signalling a GWW
  transition.}
\label{fig phase}
\end{center}
\end{figure}

Recall that $u_1= 0= {\rm Tr} W$ corresponds to an analog of the
confinement phase in gauge theory. The vanishing of all $u_n$ also has
a familiar interpretation. Let us define an eigenvalue density of the
Wilson line \eq{wilson} by
\begin{align} 
\rho(\alpha)=\frac{1}{N} \sum_{n=1}^{N} \delta(\alpha-\alpha_i).
\label{eigen value density}
\end{align}  
In terms of this, the $u_n$ are given by
\begin{align} 
u_n=\int_0^{\frac{2\pi}{\beta} } d\alpha \rho(\alpha) e^{-in \beta \alpha }.
\end{align}  
The vanishing of all $u_n$ therefore corresponds to a uniform
eigenvalue distribution. 

Let us now consider the effect of increasing the temperature, or
equivalently increasing $\bar x$. As $\bar x$ crosses $1/D$, i.e.  $T$
crosses a critical temperature given by
\begin{align}
T_{c1}=\frac{{\tilde \l}^{1/3}}{\log D}= {\l^{1/3}}
  \frac{D^{1/3}}{\log D},
\label{tc-1}
\end{align}
the sign of $a_1$ in \eq{LG} flips, while the coefficient of $|u_1|^4$
and the coefficients of all $|u_n|^2, n>1$ remains positive.  The
`$|u_n|$'s remain zero for $n>1$ \footnote{Because of couplings such
  as $u_{-2} u_1^2$ discussed in Section \ref{subsec 1/D 1d}, higher
  $u_n$'s pick up some non-zero values at higher orders in
  $1/D$. However, these can be ignored in the present discussion.}
whereas $|u_1|$ assumes a small non-zero value
\begin{align}
\vev{|u_1|} =\sqrt{ -\frac{a_1}{2b_1}} = \sqrt{\frac{(3D \log D)
    \delta T}{2{\tilde \l}^{1/3}}}-\frac{3}{8}\sqrt{6D}\left( \log D \right)^{5/2}\left(\frac{\delta T}{\tilde{\lambda}^{1/3}} \right)^{5/2}+\cdots ,
\label{u1-vev}
\end{align}
at $T= T_{c1} + \delta T$ for small and positive $\delta T$.
\footnote{$\delta T$ here is assumed to be small enough such that
  $\vev{|u_1|}$ satisfies the bound $|u_1| \le 1/2$ discussed
  below.}  

The order of the phase transition can be determined by studying the
free energy. It is easy to show that, for small $\delta T$, the Landau
Ginzburg free energy is of the form
\begin{align}
{\cal S}/(DN^2)= {\rm constant} + {\rm constant}
(\delta T)^2 \Theta(\delta T) +O\left( \frac{1}{D},
\frac1{N^2} \right).
\label{LG-saddle}
\end{align}
The second derivative of the above function with respect to $\delta T$
is discontinuous at $\delta T=0$. Thus we have a second order phase
transition. We should remark that the transition is characteristic of
the large $N$ limit and is expected to be smoothened at finite $N$, as
has been argued in \cite{Aharony:2003sx}.
 
As we increase the temperature further, we encounter another phase
transition.  To understand this phase transition, we first note that
when $u_n=0$ for all $n>1$ (which we expect to hold as long as $\bar
x$ does not cross $1/\sqrt D$), the eigenvalue density can be
represented as\footnote{By a suitable shift of the origin of the
  angle $\a$ to absorb the phase of $u_1$.}
\[
\rho(\a)=\frac\b{2\pi}\left(1+ 2|u_1|\cos(\b\a)\right).
\]
As $|u_1|$ increases from small values to $1/2$, the eigenvalue
density vanishes at $\b\a=\pi$. In the present case, the saddle point
value $\vev{|u_1|}$, \eq{u1-vev}, reaches the value $1/2$ (see
Fig. \ref{fig phase}) when $T$ equals a critical temperature 
\be
T_{c2}= T_{c1} + \frac{{\tilde \l}^{1/3}}{6D \log D}.
\label{tc-2}
\ee As $T$ crosses $T_{c2}$ we have a GWW type phase transition
\cite{Gross:1980he, Wadia:1980cp} from a gapless eigenvalue
distribution to a gapped one. The nature of this transition has been
discussed in detail in \cite{Aharony:2003sx, AlvarezGaume:2005fv},
where a Landau-Ginzburg potential of the form \eq{LG} was assumed,
with vanishing $u_n, n>1$. Our analysis above supports this
assumption, hence we can use their analysis. Following Eqn. (6.18) in
\cite{Aharony:2003sx}, we find that for temperatures just above
$T_{c2}$, $|u_1|$ grows as
\begin{align} 
|u_1| =\frac{1}{2}+\frac{\log D}{12 D^{2}}
\left(\sqrt{1+\frac{36D^3}{\tilde{\lambda}^{1/3} }(T-T_{c2}) }-1 \right)+O\left( \frac{1}{D} \right) .   
\label{u1-Tc2}
\end{align} 
By comparing this equation with the form of $|u_1|$ below $T_{c2}$
(\ref{u1-vev}), we find that the second derivative of $|u_1|^2$ (or
equivalently the third derivative of the Landau Ginzburg free energy)
with respect to temperature is discontinuous at $T_{c2}$, although the
first derivative or the value of $|u_1|^2$ is continuous. To be
precise, we find\footnote{\label{ft:sharp}The rate of change of
  $|u_1|^2$ below $T_{c2}$ is given by the expansion parameter
  $(\log D)^2 |T-T_{c2}|/{\tilde \l}^{1/3} $ while above $T_{c2}$ it is
  given by the expansion parameter $ D^3 |T-T_{c2}|/{\tilde \l}^{1/3}
  $. Hence $|u_1|^2$ changes at a much faster rate above
  $T_{c2}$.}
\begin{align}
|u_1|^2 &= \frac14 + \frac32 D \log D\, \frac{T-
  T_{c2}}{\tilde{\lambda}^{1/3} } -\frac{9D \left( \log D\right)^3
}{4} \left( \frac{T- T_{c2}}{\tilde{\lambda}^{1/3} } \right)^2 +\cdots
, \quad T \mylsim T_{c2} 
\nn 
|u_1|^2 &= \frac14 + \frac32 D \log D\,
\frac{T- T_{c2}}{\tilde{\lambda}^{1/3} } -\frac{27D^4 \log D}{2}
\left( \frac{T- T_{c2}}{\tilde{\lambda}^{1/3} } \right)^2 +\cdots ,
\quad T \mygsim T_{c2}
\label{third-order}
\end{align}
where we have ignored corrections of $O(1/D^2)$ and terms proportional
to $(T- T_{c2})^3$ and higher.  Thus, the phase transition at
$T=T_{c2}$ is third order, as in the original GWW transition.  (See
Figure. \ref{fig u1}.)

\begin{figure}
\begin{center}
\includegraphics[scale=0.75]{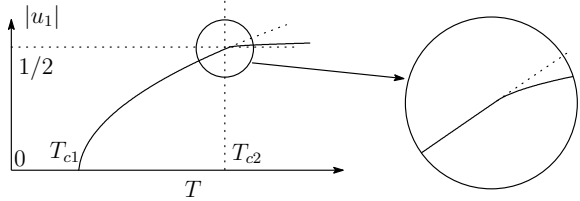}
\caption{Phase transitions: $|u_1|$ vs $T$.  As $T$ crosses $T_{c1}$,
  $|u_1|$ starts growing from zero and equals $1/2$ at $T=T_{c2}$. The
  transition at $T_{c2}$ is a third order GWW type transition. Because
  of the sharp change in the second derivative of $|u_1|$ across
  $T_{c2}$ within a very small range of temperature (see Eqn.
  \eq{third-order} and also footnote \ref{ft:sharp}), it almost
  appears like a discontinuity in the slope of the plot. However, when
  we zoom into a small temperature interval around $T_{c2}$, the $\del
  |u_1|/\del T$ is seen to be continuous, as is analytically evident
  from \eq{third-order}. Although we have plotted $|u_1|$ vs $T$ here,
  to facilitate comparison with \cite{Kawahara:2007fn}, a plot of
  $|u_1|^2$ vs $T$ shows exactly similar features.}
\label{fig u1}
\end{center}
\end{figure}

Beyond $T_{c2}$ it is in principle possible to have further phase
transitions to multiple-gap phases. In case of unitary matrix models
with general single trace actions of the form $\sum_n c_n u_n + {\rm
  c.c.}$ it was shown in \cite{Jurkiewicz:1982iz, Mandal:1989ry} that
additional gaps can open up in the eigenvalue distribution as the
temperature is varied.  We will not attempt to study this issue in
this paper, since the analysis in the relevant ranges of temperatures
is complicated.  The numerical analysis of \cite{Kawahara:2007fn},
appears to suggest, however, that the only phase transitions in the
model are the two already discussed above.

\paragraph{High temperature:}

Once the temperature becomes very high, $\beta_{\rm eff} \ll 1$, the model
again admits analytic treatment. In this region, the eigenvalue
density approaches a delta function and we can approximate $u_n=1$.
The saddle point equation (\ref{d=1 saddle point equation}) then
becomes
\begin{align} 
-\frac{\triangle^3}{2 \tilde{\lambda} }+
\frac{1}{2}+\frac{e^{-\beta\triangle}}{1-e^{-\beta\triangle}}=0.
\end{align} 
We can approximately solve it for small $\beta$ as, 
\begin{align} 
\triangle=\left( \frac{2\tilde{\lambda} }{\beta} \right)^{1/4}.
\end{align} 
This result is consistent with the $d=0$ condensate in
(\ref{condensation 0d}) by identifying $\tilde \l|_{d=0}$ as
$\tilde{\lambda}/\beta $.\\

\ni We end this subsection with a few comments:

%\begin{quote}

\paragraph{(a) Small $|u_1|$ approximation:} 
Since the second phase transition happens when $|u_1|$ reaches 1/2,
one may worry about the validity of small $u_n$ approximation in
(\ref{saddle-value-d=1}). There is no problem, however, since terms
involving $u_n$ in the saddle point equation (\ref{d=1 saddle point
  equation}) are also suppressed by $x^n \sim 1/D^n$ around the
critical temperature.  Therefore, even if $|u_1|$ is not small, our
analysis is valid.

\paragraph{(b) Large $D$ limit vs $1/D$ expansion:} 

It is clear from the phase transition temperatures \eq{tc-1} and
\eq{tc-2} that the phase transitions disappear in the large $D$ limit. Hence we
should simply regard $D$ as large but not take the strict $D
\to\infty$ limit if we want to explore criticality.
 
%\end{quote}

\subsection{ $1/D$ correction to the effective action}
\label{subsec 1/D 1d}

In the previous section, we have considered the effective theory
(\ref{effective action d=1}) including the $1/D$ term \eq{FP} from the
gauge fixing and discussed the phase transition.  However, in addition
to this $1/D$ term, other $1/D$ corrections can arise from $S_{int}$
in \eq{action 1dMM}.  It corresponds to ${\cal S}_{1,0}$ in
(\ref{double-expansion}) in the 0 dimensional model.  Hence we have to
evaluate them and show that these corrections are sub-dominant around
the critical point.

The terms we should look for at this order, in so far as the issue of
phase transitions is concerned, are as follows. Besides the explicit
corrections to $|u_1|^2$ and $|u_1|^4$ in (\ref{LG}), the corrections
to the gauge-field independent terms are also relevant, since they
contribute to the saddle point equation (\ref{d=1 saddle point
  equation}).  Interaction terms like $u_1^2 u_{-2}$ also affect the
Landau-Ginzburg type potential (\ref{LG}) by generating an effective
$|u_1|^4$ term.  However as we show below Eqn. (\ref{two-loop
  interaction}), the corrections to the coefficient of $|u_1|^4$ from
these interactions are order $1/D^4$ and we ignore them
here\footnote{Although the cubic interaction $u_1^2 u_{-2}$ merely
renormalizes the coefficient of the $|u_1|^4$ term in the
   Landau-Ginzburg potential (\ref{LG}), one needs to be careful
about integrating out the $u_2$ consistent with the positivity
constraint of $\rho(\a)$.}.  Thus the
relevant terms in the effective action are
\begin{align} 
{\cal S}(\triangle, \{u_n\})/(DN^2)=C_0+C_2 |u_1|^2 +C_4 |u_1|^4
+\cdots ,
\label{s-eff-all}
\end{align} 
and we can explicitly calculate them,
\begin{align} 
\kern-100ptC_{0}=&-\frac{\beta \triangle^4}{8 \tilde{\lambda}
}+\frac{\beta \triangle}{2} \nonumber 
\\ &+\frac{\beta \triangle}{D}
\left[ \left( 1+\frac{\tilde{\lambda} }{4\triangle^3}
  \right)^{\frac{1}{2} } -1-\left(\frac{\tilde{\lambda}
  }{4\triangle^3} \right)-\frac{1}{4}\left(\frac{\tilde{\lambda}
  }{4\triangle^3} \right)^2 \right], 
\end{align}
\gap{-1}
\begin{align}
&\kern-75ptC_{2}=
\left(\frac{1}{D}-x \right) +\frac{\beta \triangle}{D} \Biggl[
  \left(\frac{\tilde{\lambda} }{4\triangle^3} \right) \left(
  1+\frac{\tilde{\lambda} }{4\triangle^3} \right)^{-\frac{1}{2} }
  \nonumber 
\\ &\kern-50pt+\frac{\frac{\tilde{\lambda}
    }{4\triangle^3}}{1+\frac{\tilde{\lambda} }{4\triangle^3} }
  -4\left(\frac{\tilde{\lambda} }{4\triangle^3}
  \right)-3\left(\frac{\tilde{\lambda} }{4\triangle^3} \right)^2
  \Biggr]x +O(x^2),
\end{align}
\gap{-1}
\begin{align}
C_{4}=& \frac{\beta
  \triangle}{4D} \left[ -\left(\frac{\tilde{\lambda} }{4\triangle^3}
  \right)^2 \left( 1+\frac{\tilde{\lambda} }{4\triangle^3}
  \right)^{-\frac{3}{2} } -2\left(\frac{\tilde{\lambda}
  }{4\triangle^3} \right)^2 \right]x^2 \nonumber 
\\ &
+\frac{\beta \triangle}{2D}(2+\beta \triangle ) \left[
  -\frac{\left(\frac{\tilde{\lambda} }{4\triangle^3} \right)^2}{
    \left( 1+\frac{\tilde{\lambda} }{4\triangle^3} \right)^2 }
  -2\left(\frac{\tilde{\lambda} }{4\triangle^3} \right)^2 \right]x^2
+O(x^3).
\end{align} 
Here we have omitted higher $x$ terms, since we are interested in a
range of temperatures below or around the critical temperature $x \sim
1/D$ \footnote{Note that the small $x$ expansion is valid for low
  temperatures.  It means that our analysis works well for large
  effective coupling $\lambda_{\rm eff}$ (\ref{lam-eff}) (as long as
  it does not scale with $D$)\label{footnote strong coupling}.
This assumption is in particular valid around the phase transitions.}
details of the derivation are shown in Appendix \ref{app all loop}.

As in \eq{saddle-value-d=1}, we solve the saddle point
equation for $\triangle$ in powers of $u_1$, to
obtain:
\begin{align} 
\frac{ \triangle}{\tilde{\lambda}^{1/3} }=1+\frac{1}{D}
\left(\frac{7\sqrt{5}}{30}-\frac{9}{32}  
 \right)  +  \frac{2}{3}\bar{x}  |u_1|^2 +\cdots .
 \label{condensation d=1 1/D}
\end{align} 
Here the $|u_1|^4$ and higher order terms do not affect \eq{LG'} and are
dropped.  Substituting this in \eq{s-eff-all}, we get
\begin{align}
{\cal S}/(DN^2)= \beta \tilde{\lambda}^{1/3} 
\epsilon_0 +  a_1' |u_1|^2 + b_1'|u_1|^4+\cdots,
\label{LG'}
\end{align}
with
\begin{align} 
\epsilon_0=&\frac{3}{8}+\frac{1}{D}
\left(-\frac{81}{64}+\frac{\sqrt{5}}{2}   \right), \\  
a_1'=&\frac{1}{D}-\bar{x}- \frac{\tilde{\lambda}^{1/3}\beta }{D} \left(
\frac{203}{160} -\frac{\sqrt{5}}{3} \right) \bar{x},
\\ b_1'=&\frac{\tilde{\lambda}^{1/3} \beta }{3}
\bar{x}^2-\frac{\tilde{\lambda}^{1/3} \beta }{D} \left(
\tilde{\lambda}^{1/3} \beta \left( \frac{2\sqrt{5}}{9}
-\frac{229}{300} \right) +\frac{391\sqrt{5}}{1800}-\frac{3181}{2400}
\right) \bar{x}^2.
\label{b 1/D}
\end{align} 
It is obvious that these equations constitute $O(1/D)$ fractional
corrections to various quantities appearing in \eq{LG}. 

As argued in the previous subsection, the phase transition temperature
$T_{c1}$ is characterized by vanishing of $a_1'$ and $T_{c2}$ is given
by $b_1'= -2 a_1'$. This gives us the following corrected values of
the transition temperatures:
\begin{align} 
\beta_{c1} \tilde{\lambda}^{1/3} &  = \log D
\left( 1 +\frac{1}{D} \left( \frac{203}{160} -\frac{\sqrt{5}}{3}
\right) \right) .
\label{tc1}\\
\beta_{c2} \tilde{\lambda}^{1/3}
&-\beta_{c1} \tilde{\lambda}^{1/3} \nonumber \\
 =&\frac{\log D}{D} \Biggl[ -\frac{1}{6}    
 +\frac{1}{D}
 \left( 
 \left( -\frac{499073}{460800}+\frac{203\sqrt{5}}{480}   
 \right) \log D 
 -\frac{1127\sqrt{5} }{1800}+\frac{85051}{76800} 
  \right) 
  \Biggr].
  \label{tc2}
\end{align} 
Although the analysis in this subsection leads to subleading
corrections to the phase transition temperatures, it is easy to see
that the nature of the phase transitions derived in the previous
subsection remains unaltered.

\subsection{$1/D$ expansion vs. numerical calculation}
\label{sec 1/9}

In this section, we evaluate the critical temperatures and a few other
quantities, using results in the previous subsections in the large $D$
expansion, and compare them with the numerical results for $D=9$,
which were studied in \cite{Aharony:2004ig, Aharony:2005ew,
  Kawahara:2007fn, Azeyanagi:2009zf}.

The $d=1$ model was numerically analyzed in \cite{Aharony:2004ig,
  Aharony:2005ew} (see Section \ref{sec:review} for connection to D
branes) where it was suggested that the system undergoes a weakly
first order Gross-Witten-Wadia type transition (characterized by the
development of a gap in the eigenvalue distribution $\rho(\a)$). A
more detailed numerical study \cite{Kawahara:2007fn} subsequently
claimed that in stead of a single first order transition, it consists
of two higher order phase transitions: (a) from uniform to non-uniform
$\rho(\a)$, followed closely by (b) a GWW type transition in which a
gap appears. This is in agreement with the picture of the two
transitions we derived in the previous subsections (see
Figure \ref{fig phase diagram}).  Let us compare between our results and those of \cite{Kawahara:2007fn} in some detail.

We first compare the two critical temperatures derived from the
numerical analysis in \cite{Kawahara:2007fn} with our large $D$
expansion.  In order to do it, we note that the dimensionless
temperature defined in \cite{Aharony:2004ig, Aharony:2005ew,
  Kawahara:2007fn, Azeyanagi:2009zf} and \eq{beta-eff} is
\begin{align}
T_{\rm eff} \equiv \frac{1}{\beta_{\rm eff}} = \frac1{ \lambda^{1/3}
  \beta}= \frac{D^{1/3}}{\tilde{\lambda}^{1/3} \beta}.
\end{align} 
In the units $\lambda=g^2N=1$, used by \cite{Kawahara:2007fn}, $T_{\rm eff}$ is
simply $T$. By employing the same unit, we obtain the critical
temperatures as in Table \ref{compare with num}.  The leading order
results in the $1/D$ expansion are from Eqns. (\ref{tc-1}) and
(\ref{tc-2}), and the next order is from Eqns. (\ref{tc1}) and
(\ref{tc2}).

Similarly, we can also compare the value of the condensate
$\triangle^2$ and the free energy in the confinement phase, which are
given by
\begin{align} 
R^2 \equiv \frac{g^2}{N}  
\langle \Tr Y^I Y^I \rangle |_{\beta \rightarrow \infty} =  
\frac{1}{2} \triangle^2|_{\beta \rightarrow \infty},  \quad
 F_0 \equiv -\frac{1}{\beta N^2} \log Z |_{\beta \rightarrow \infty}.
 \label{condensation and free energy}
\end{align} 
Those can be derived from (\ref{saddle-value-d=1}) and (\ref{LG}) in
the leading order, and (\ref{condensation d=1 1/D}) and (\ref{LG'}) in
the next order.  The results are also summarized in table \ref{compare
  with num} and our results agree with the numerical result remarkably
well\footnote{Note that we call the critical temperature from uniform
  to non-uniform distribution as $T_{c1}$ and the next GWW type as
  $T_{c2}$. However, in \cite{Kawahara:2007fn}, they used the opposite
  notation.  $R^2$ and $F_0$ in (\ref{condensation and free energy})
  are defined as $r_0^2$ and $\epsilon_0$ in \cite{Kawahara:2007fn}.}.

\begin{table}
\begin{center}
\begin{tabular}{l|ccccc}
\hline
&$T_{ c1}$&$T_{ c2}$&$R^2$&$ F_0$ \\ 
\hline
Numerical result & 0.8761 & 0.905  & 2.291 & 6.695 \\ 
Leading large $D$ result & 0.947 & 0.964  & 2.16 & 7.02 \\
Large $D$ including $1/D$ effect & 0.895 & 0.917  & 2.28 & 6.72  \\
\hline
\end{tabular}
\caption{Comparison with numerical results derived in
  \cite{Kawahara:2007fn} for $D=9$ and our large $D$ analysis.  Here
  we have used $\lambda=g^2N=1$ units.  We list the critical
  temperature $T_{c1}$ and $T_{c2}$, and the condensation and the free
  energy at the confinement phase defined in (\ref{condensation and
    free energy}).  The first line is the numerical result.  The
  values in the second line are the leading large $D$ results.  The
  third line is the result including the first $1/D$ correction. The
  fractional differences from the numerical results can be checked to
  be order $1/D$ in the second line and $1/D^2$ in the third line, as
  expected.}
\label{compare with num}
\end{center}
 \end{table}

We note here that although we find excellent quantitative agreement
with \cite{Kawahara:2007fn}, the more qualitative inferences in
\cite{Kawahara:2007fn} regarding the order of the phase transitions
are different from ours. The phase transitions at $T_{c1}$ and
$T_{c2}$ are claimed in \cite{Kawahara:2007fn} to be of 3rd order and
2nd order, respectively, while in our analysis they are the other way
around. We believe that this difference may be due to the fact that in
numerical work it is not easy to ascertain the order of a transition
except when it is a strong first order transition. The transition at
$T_{c1}$ in our analysis is described by a classic Landau Ginzburg
potential which describes a second order transition.  For a LG
potential involving only $u_1$ to describe a third order transition as
suggested in \cite{Kawahara:2007fn}, we need a $u_1^3$ term which is
disallowed by the symmetries of the theory.  Likewise, a second order
transition at $T_{c2}$ is inferred in \cite{Kawahara:2007fn} by noting
a jump in $\partial |u_1|/ \partial T$. We find, on the other hand,
that there is a sharp, but continuous change in this quantity (see
Figure \ref{fig u1} and Eqn. \eq{third-order}). We note that our
analysis, of course, ignores corrections of order $1/D^2$; however, we
do not expect any qualitative changes in the above conclusions for
large values of $D$ such as $D=9$\footnote{\label{ftnt:nishimura}
  J.~Nishimura informed us that he agrees with the conclusion obtained
  in this paper and that the numerical data in \cite{Kawahara:2007fn}
  around $T_{c1}$ are also consistent with a second order phase
  transition.  He also mentioned that fitting the data with that
  assumption leads to a slight increase in their estimate on $T_{c1}$,
  which further improves the agreement with our value of $T_{c1}$.
  The estimate of $T_{c2}$ in \cite{Kawahara:2007fn}, on the other
  hand, does not depend on the assumed order of transition in their
  analysis.  Thus, the agreement with our value of $T_{c2}$ is not
  affected.  We thank J.~Nishimura for providing us with the results
  of this reanalysis.}.

\paragraph{$1/D$ expansion for small $D$:}

Ref. \cite{Azeyanagi:2009zf} also numerically analysed the transition
for $D=2$ and 3, and found two transitions as in the $D=9$ case in
\cite{Kawahara:2007fn}.  In our
study too, we have two transitions for $D=2$ and 3, since $b_1'$ in
(\ref{b 1/D}) is positive even for these values of $D$. We summarise
these results as follows:
\begin{center}
\begin{tabular}{l|cccc}
\hline 
 & $ T_{ c1} $ ($D=2$) & $ T_{ c2}$ ($D=2$)& $ T_{ c1} $ ($D=3$) & $ T_{ c2}$ 
($D=3$)  \\ \hline 
Our result & 1.4 & 1.6 & 1.1 & 1.2 \\
Numerical result  & 1.12 & 1.3 & 0.93 & 1.1\\  \hline
\end{tabular}
\end{center}
% \end{table}
The critical temperatures in their numerical results are close to our
results \footnote{Their interpretation of the order of the phase
  transitions is the same as in \cite{Kawahara:2007fn}, and is
  different from ours. The explanation of this discrepancy is similar
  to the $D=9$ case as mentioned above.}. Our phase transition
temperatures also show agreement with the critical temperature
numerically evaluated in Ref. \cite{Aharony:2005ew} for $D=4$.

These results suggest that our analysis seems to work even for such
small values of $D$.  However, we do not have a detailed understanding
of such an unexpected agreement.

\section{D brane interpretation}\label{sec:review}

The $d=1$ model \eq{matrixqm-action} appears in various contexts, as
mentioned in the Introduction. The context closest to the contents of
this paper is that of \cite{Aharony:2004ig, Aharony:2005ew,
  Kawahara:2007fn}, which we briefly review in this section.

Let us consider thermal D0 branes in $R^8 \times S^1$.  The
distribution of the branes is dynamically determined and for a certain
parameter region, the geometry becomes a black string winding around
the $S^1$.  If we increase the radius of the $S^1$ beyond a critical
radius, the Gregory-Laflamme instability mode \cite{Gregory:1993vy}
appears and a black hole solution, localized on the $S^1$, is
favoured.  It is argued in \cite{Aharony:2004ig} (see also
\cite{Kol:2004ww}) that this black string/black hole transition is
first order.

\begin{figure}
\begin{center}
\includegraphics[scale=0.75]{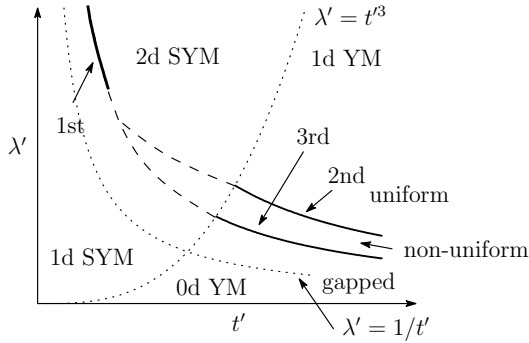}
\caption{Phase diagram of the $d=2$ SYM (\ref{d=2}).  Below
  $\l'={t'}^3$, the temporal KK modes can be ignored and below
  $\l'=1/t'$, the spatial KK modes can be ignored.  Thus the effective
  $d=1$ (bosonic) YM description is valid on the right of the
curve $\l'= {t'}^3$. The overlap of this region with the
region below  $\l'=1/t'$ additionally 
  admits an effective $d=0$ description.  The two
  phase transition lines below $\l'={t'}^3$ are given by $\l't'
  =1/T_{c1}^3$ and $\l't' =1/T_{c2}^3$, where $T_{c1,c2}$ are given in
  \eq{tc1} and \eq{tc2}.  A similar phase structure was earlier
  inferred in \cite{Kawahara:2007fn} on the basis of numerical
  analysis. }
\label{fig phase diagram}
\end{center}
\end{figure}
Through gauge/gravity duality \cite{Itzhaki:1998dd}, we expect this
transition to be reproduced by a $d=1$ thermal SYM with a compact
adjoint scalar at strong 'tHooft coupling.  By using a
T-duality \cite{Taylor:1996ik}, this model can be mapped to a 2d SYM on
$T^2$ 
\begin{align} 
S=&\frac1{g_2^2}  \int_0^L \kern-8pt dx 
\int_0^{\b_2} \kern-10pt  dt\  
\Tr \left(\frac14 F_{\mu\nu}^2 + \frac12 \sum_{I=1}^{8} 
D_\mu Y_I D^\mu Y_I - \frac14 \sum_{I,J} [Y_I, Y_J]^2 \right) 
+ {\rm fermions}.
\label{d=2}
\end{align} 
This theory is characterized by two independent dimensionless
constants: (a) $\l'=\l_2 L^2$ where $\l_2 = g_2^2 N $ is the 'tHooft
coupling, and (b) $t'= L/\b_2$, the dimensionless temperature.
However, the analysis of this theory at strong 'tHooft coupling is
difficult.  Instead of investigating the above transition at strong
coupling, the gauge theory allows us to study a continuation of the
phase transition to weak 'tHooft coupling.

It has been argued in \cite{Aharony:2004ig, Aharony:2005ew} that in
the range of temperatures given by ${\l'}^{1/3} < t'< 1/\l'$ all
fermionic modes as well as both the spatial and temporal KK modes can
be ignored.  The theory is then governed by just the zero modes which
describe the $d=0$ model studied in Section \ref{sec 0d MM}. (See
Figure \ref{fig phase diagram}).  As $\l' t'$ grows to order unity,
the spatial KK modes cannot be ignored any more, though the temporal
KK modes can still be ignored. In fact, in the range of temperatures
and coupling ${t'}^3> \l'$, the theory \eq{d=2} reduces to the $d=1$
model (\ref{matrixqm-action}) with the spatial circle of length $L$
identified with $\b$ of \eq{matrixqm-action}.  The $d=1$ 'tHooft
coupling $\l$ is identified with $\l_2/\b'$ so that $\l_{\rm eff}$
appearing in \eq{lam-eff} is identified as
\begin{align} 
\l_{\rm eff}= \l_2 L^3/\b_2 = \l't' . 
\label{effective coupling from d2}
\end{align} 
Note that the transitions in the $d=1$ model, which we studied in
Section \ref{sec d=1 contd}, happen around $\lambda_{\rm eff} =
1/T_{\rm eff}^3 \approx 1.4$ from Table \ref{compare with num}.  These
transitions can indeed be regarded as transitions in the $d=2$ model
(\ref{d=2}) if $t'^4> \lambda_{\rm eff} $.  Thus we can reliably
expect these transitions to be the continuation of the black
string/black hole transition to weak 'tHooft coupling\footnote{The
  ``weak'' coupling here refers to the $d=2$ coupling $\l'$ which
  satisfies $\l'<{t'}^3$ (see Figure \ref{fig phase diagram}). We
  should remark that the analysis in this paper is valid even for
  large values of the $d=1$ 'tHooft coupling $\l_{\rm eff}$ as we
  explained in footnote \ref{footnote strong coupling}. The
  equivalence with the $d=2$ model, however, is valid only for
  temperatures $t'^4> \lambda_{\rm eff}$.}.

As has been suggested first in \cite{Susskind:1997dr}, the eigenvalue
distribution of the Wilson loop (\ref{eigen value density}) is related
to the geometry of the D0 branes.  A uniform (non-uniform) gapless
eigenvalue distribution corresponds to a uniform (non-uniform) black
string winding around the $S^1$, whereas a gapped distribution
corresponds to a black hole localized on the $S^1$. Now, we found in
Section \ref{sec d=1 contd} that the uniform eigenvalue distribution
is favoured at low temperature and a gapped distribution is favoured
at high temperature while a non-uniform distribution exists between
those two phases.  Since the temperature in the $d=1$ model is mapped
to the radius of the original $S^1$, there should be a phase
transition from a black string to a black hole as the radius of the
circle is increased, which indeed is the case.  The fact that our
transition consists of two closely spaced transitions disagrees,
however, with the single first order transition in the gravity
description.  A plausible resolution is as follows. It is easy to see
that if $b_1$ in (\ref{LG}) is negative when $a_1$ vanishes, there is
only one, first order, phase transition instead of the two transitions
\cite{Aharony:2003sx}.  Therefore the gravity picture can be
reconciled with the gauge theory calculation if the sign of $b_1$ in
(\ref{LG}) flips at some higher value of coupling in the two
dimensional model. At such a value the two phase transition lines
found at weak coupling will merge and yield a single first order
transition line (see Figure \ref{fig phase diagram}).

\section{Conclusion}\label{sec conclusion}

In this paper we have developed a technique of solving matrix models
($d=0,1$) which are dimensional reductions of $D+d$ dimensional
bosonic YM theory to $d$ dimensions. The technique involves working in
a $1/D$ expansion, which allows us to analytically compute free
energies and other thermodynamic quantities. In the $d=1$ case our
results show that the system undergoes a double phase transition: a
second order phase transition which signals onset of a non-uniformity
of the eigenvalue distribution $\rho(\a)$ of the Wilson line, followed
by a third order GWW phase transition which signals development of a
gap in $\rho(\a)$. Following the arguments in \cite{Aharony:2004ig,
  Aharony:2005ew, Kawahara:2007fn, Azeyanagi:2009zf}, we interpreted this double
transition as a continuation of the Gregory-Laflamme (black
string/black hole) phase transition to weak coupling. Our results
agree with the numerical results of \cite{Kawahara:2007fn}, and offers
an analytic resolution of the issue of the order of the phase
transitions.

The large $D$ technique developed in this paper is in principle
applicable to a variety of bosonic matrix models involving
commutator-squared interactions. The applicability of our techniques
would be greatly enhanced if we are able to extend them to higher
dimensional models $d>1$ and to include supersymmetry.  The extension
to higher dimensional models appears to have the technical hurdle of
computing $\log {\rm det} (D_\mu^2 + iB_0)$ with dynamical gauge fields
without making a coupling constant expansion. One possibility is to
find regions of parameter space in which an effective $d=1$
description arises (as in the $d=2$ toroidal model described in
Section \ref{sec:review}) and work around that limit. The
supersymmetric extension of the large $D$ methods appears more
challenging, even qualitatively, since the number of bosons and
fermions grow at different rates as $D$ grows large. We hope
to come back to some of these issues in a future publication.

In this paper we have been concerned with thermodynamic properties of
the matrix models. Another possible application of our methods could
be to address dynamical questions. Indeed, one of the motivations of
this paper was to apply these techniques to derive an effective action
for gauge fields in the time-dependent context and to study dynamical
phase transitions using this effective action. Work in this direction
is in progress \cite{progress}.

\subsection*{Acknowledgement} 

 We would like to thank Spenta Wadia for collaboration in an ongoing
 work on dynamical black hole/black string transitions \cite{progress}
 which inspired the present paper, and for sharing numerous insights.
 We would like to thank Adel Awad, Avinash Dhar, Sumit Das, Ian
 Ellwood, Barak Kol, Oleg Lunin, Samir Mathur, Jun Nishimura, Toby
 Wiseman and especially Shiraz Minwalla for useful discussions.  We
 would like to thank Jun Nishimura for sharing with us the result a
 reanalysis of \cite{Kawahara:2007fn} which further improves the
 agreement with our analysis.  T.M. would like to thank the theory
 group at KEK for their kind hospitality, where part of this work was
 done. G.M. would like to thank the organizers of the Benasque
 conference on Gravity (July 2009), the organizers of the QTS6 meeting
 in Lexington, the University of Kentucky, Lexington and the School of
 Natural Sciences, IAS, Princeton for hospitality during part of this
 project.

\appendix
\section{Some results involving $M_{ab,cd}$}
\label{app Mabcd}

In this section, we will calculate several quantities involving
$M_{ab,cd}$, for example, $\Tr M^{n}$, $M^{-1}$, the eigenvalues of
$M$ etc. which are important in solving both the $d=0$ and the $d=1$
models. We begin by first investigating algebraic properties of
$M_{ab,cd}$.

\subsection{Algebraic properties of  $M_{ab,cd}$}

As in (\ref{def Mabcd}), $M_{ab,cd}$ is defined as
\begin{align} 
M_{ab,cd} = -\frac{1}{4} \Bigl\{ \Tr[\l_a,
  \l_c][\l_b, \l_d] +(a\leftrightarrow b)+(c\leftrightarrow
d)+(a\leftrightarrow b,c\leftrightarrow d) \Bigr\}.
\label{def M-app}
\end{align}
$M_{ab,cd}$ has four adjoint indices and is symmetric under the
interchanges $a\leftrightarrow b$ and $c \leftrightarrow d$.  Hence,
in the $SU(N)$ case, we can regard $M_{ab,cd}$ as an $N^2(N^2-1)/2
\times N^2(N^2-1)/2$ matrix by identifying $ab$ and $cd$ as two single
indices.  Equivalently, we can regard $M$ as an operator acting on the
$N^2(N^2-1)/2$ dimensional vector space $V_B$ (whose elements can be
regarded as $B_{ab}$) labeled by a symmetric pair of adjoint indices
$ab$, on to the same vector space ({\em i.e} $M$ is an endomorphism of
the vector space $V_B$).

\begin{figure}
\begin{center}
\includegraphics[scale=1.5]{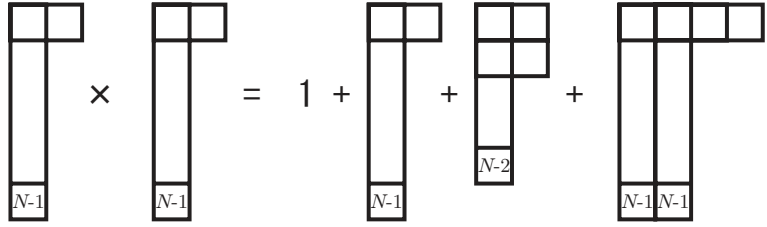}
\caption{Irreducible decomposition of the $N^2(N^2-1)/2$ dimensional
  vector space $V_B$ labelled by a symmetric pair of adjoint indices
  $ab$, regarded as a symmetric product of `adjoint' $\times$
  `adjoint'.  The dimensions of the representations in the RHS are
  $ 1$, $N^2-1$, $N^2(N+1)(N-3)/4$ and $N^2(N-1)(N+3)/4$ respectively.
  These correspond to the space obtained from the projection through
  $K_0$, $K_1$, $K_2$ and $K_3$.  Note that $SU(2)$ and $SU(3)$ are
  exceptional.  In $SU(2)$ the second and the third representations
  are absent and in $SU(3)$ the third one is absent
  \cite{Slansky:1981yr}.}
\label{fig young}
\end{center}
\end{figure}
To proceed, let us decompose this $N^2(N^2-1)/2$ dimensional vector
space $V_B$ into irreducible representations of $SU(N)$.  The
decomposition is shown in Figure \ref{fig young} and we obtain four
irreducible representations.  Let us define a vector in this space as
$B_{ab}$, then the first ${\bf 1}$ in Figure \ref{fig young} is the `trace'
part $B_{aa}$ \footnote{Here `trace' of a vector $B_{ab}$ means the
  sum of the two adjoint indices. e.g. $B_{aa}$.}  and the latter
three constitute irreducible decomposition of the symmetric
`traceless' vector. Correspondingly we can define four projection
operators $K_{iab,cd}$ ($i=0,1,2,3$) acting on this vector space such
that $K_{i ab,cd} B_{cd}$ belongs to the $i$-th irreducible
representation,
\begin{align} 
B_{ab}=(K_0 B)_{ab}+(K_1 B)_{ab}+(K_2 B)_{ab}+(K_3 B)_{ab} \nonumber .
\end{align} 
We will show that the endomorphism $M_{ab,cd}$ can also be decomposed
following the above equation and hence can be represented as a linear
combination of the $K_i$ (see \eq{M and K}).

It is possible to construct the $K_i$ explicitly. Let us define the
following four matrices \cite{Hotta:1998en}:
\begin{align} 
F_{ab,cd}=& \frac{1}{4}\left[ {\rm Tr}\left(\lambda^a \lambda^b
\lambda^c \lambda^d \right) +(a\leftrightarrow b) +(c\leftrightarrow
d) +(a\leftrightarrow b,~c \leftrightarrow d) \right], \nonumber
\\ G_{ab,cd}=&
\frac{1}{2}\left[ {\rm Tr}\left(\lambda^a \lambda^c \lambda^b
\lambda^d \right) +(a\leftrightarrow b) \right] ,\nonumber
\\ H_{ab,cd}=&
\delta_{ab} \delta_{cd}, \nonumber
\\ I_{ab,cd}=& \frac{1}{2} \left( \delta_{ac}
\delta_{bd}+\delta_{ad}\delta_{bc} \right) .
\label{def Iabcd}
\end{align} 
Those matrices satisfy the following relations,
\begin{align} 
F^2=\frac{1}{2}\left(1+\frac{2}{N^2} \right)H +
\frac{N}{2}\left(1-\frac{4}{N^2} \right)F,~~ FG=-\frac{2}{N}F
+\frac{1}{N^2}H, \nonumber 
\\ FH=N(1-\frac{1}{N^2})H,~~G^2=I+\frac{1}{N^2}H-\frac{2}{N}F,~~GH
=-\frac{1}{N}H, \nonumber
\\ H^2=(N^2-1)H,~~FI=F,~~GI=G,~~HI=H~~,I^2=I
\label{product FGHI}
\end{align} 
where the product $AB $ is defined by $(AB)_{ab,ef}=
A_{ab,cd}B_{cd,ef}$.  Because of the cyclic property of the trace,
this product satisfies $AB=BA$.  Note that $I_{ab,cd}$ plays the role
of the identity matrix. In order to derive these relations, we have
employed the identity
\begin{align} 
\sum_{a=1}^{N^2-1} \lambda^a_{ij}\lambda^a_{kl} =
\delta_{il}\delta_{jk}-\frac{1}{N}\delta_{ij}\delta_{kl},
\end{align} 
where $i,j$ are fundamental indices.

By using these matrices, the four projection operators are represented as:
\begin{align} 
K_0 &\equiv \frac{1}{N^2-1}H, \qquad K_1 \equiv \frac{2N}{N^2-4}
\left( F-\frac{1}{N}H  \right) , \nonumber \\ 
K_2&\equiv \frac{1}{2} \left(I-G-\frac{2}{N-2}\left(F-\frac{1}{N}H  
\right)-\frac{1}{N(N-1)}H    \right),  \nonumber \\
K_3&\equiv \frac{1}{2} \left(I+G-\frac{2}{N+2}\left(F-\frac{1}{N}H  
\right)-\frac{1}{N(N+1)}H    \right) . 
\label{def K}
\end{align} 
We can show that these matrices satisfy,
\begin{align} 
K_{i} K_{j} = \delta_{ij} K_{i},
\label{product KK}
\end{align} 
and 
\begin{align} 
I_{ab,cd}= K_{0ab,cd}+ K_{1ab,cd} +  K_{2ab,cd} + K_{3ab,cd}.
\label{I and K}
\end{align} 
Thus the $K_i$ are indeed projection operators.  Actually, we can find
the correspondence between $K_i$ defined in (\ref{def K}) and the
irreducible representations in Figure \ref{fig young}.  For example,
$K_{0ab,cd}$ acts on a vector $B_{cd}$ as
\begin{align} 
K_{0ab,cd} B_{cd}=
\frac{1}{N^2-1} \delta_{ab} B_{cc}.
\label{projection K0}
\end{align} 
Thus $K_0$ maps the vector to the `trace' $B_{cc}$.  Therefore $K_0$ is the
projection operator corresponding to {\bf 1} in the RHS of Figure \ref{fig
  young}.  Similarly we can find the correspondence for other $K_i$.

In addition to such explicit identifications, we can calculate the
traces of $K_i$ as\footnote{Here `trace' of an endomorphism matrix
  $A_{ab,cd}$ is defined as ${\rm Tr} A\equiv A_{ab,ab}$}
\begin{align} 
K_{0ab,ab}&=1,\quad
K_{1ab,ab}=N^2-1 ,\nonumber \\
 K_{2ab,ab}&=\frac{N^2(N+1)(N-3)}{4}  , \quad
K_{3ab,ab}=\frac{N^2(N-1)(N+3)}{4}.
\label{trace K}
\end{align} 
The values of the traces are equivalent to the dimensions of the
irreducible representations shown in Figure \ref{fig young}.  This is
another evidence for the correspondence between $K_i$ and those
representations.

By using (\ref{def M-app}), (\ref{def Iabcd}) and (\ref{def K}),
$M_{ab,cd}$ can be described as
\begin{align} 
M_{ab,cd}&=2(F_{ab,cd}-G_{ab,cd}) \nonumber \\
&=2N K_{0ab,cd}+N K_{1ab,cd} + 2K_{2ab,cd} -2 K_{3ab,cd} .
\label{M and K}
\end{align}

\subsection{Results involving  $M_{ab,cd}$}
In this subsection we use the tools developed above to calculate
several quantities associated with $M_{ab,cd}$ necessary in the study
of the matrix models in this paper.

\paragraph{The inverse $M^{-1}$:}

Equation (\ref{gauss-trick}) assumes the existence of $M^{-1}$.  From
(\ref{I and K}) and (\ref{M and K}), we explicitly obtain 
\be
M^{-1}_{ab,cd} =\frac{1}{2N} K_{0ab,cd}+\frac{1}{N} K_{1ab,cd} +
\frac12 K_{2ab,cd} -\frac12 K_{3ab,cd} .
\label{m-inverse}
\ee
By using this, we can calculate $M^{-1}_{ab,cc}$, which is necessary to
derive (\ref{condensation Y}) and (\ref{action 1dMM}).  First we can
show $K_{iab,cc}=0$ for $i\ne 0$, since the irreducible
representations in Figure \ref{fig young} are `traceless' except {\bf 1}.
Then we obtain 
\be M^{-1}_{ab,cc} =\frac{1}{2N}
K_{0ab,cc}=\frac{1}{2N}\delta_{ab}.  
\label{m-inv-tr}
\ee

\paragraph{Eigenvalue of $M_{ab,cd}$:}
We now derive the eigenvector and eigenvalue of $M_{ab,cd}$.  By using
(\ref{M and K}), we obtain the eigenvector as
\begin{align} 
& M_{ab,cd}\left\{ (K_0B)_{cd}+(K_1B)_{cd}+(K_2B)_{cd}+(K_3B)_{cd}
  \right\} \nonumber \\ = &2N
  (K_0B)_{ab}+N(K_1B)_{ab}+2(K_2B)_{ab}-2(K_3B)_{ab}.
\label{eigen M}
\end{align} 
where $B_{ab}$ is a general vector.  Note that the eigenvalue of
$(K_3B)_{ab}$ is negative, which makes the action (\ref{gauss-trick})
not positive definite. As we have remarked in footnote \ref{footnote N}, this can be
dealt with by making appropriate choices for integration contours for
different elements of $B_{ab}$.

\paragraph{Calculation of $\Tr M^{n}$:}
We now calculate ${\rm Tr} M^n=M^n_{ab,ab}$, which appears in the loop
calculation of the matrix model.  By using Eqns. (\ref{product KK}) and
(\ref{M and K}), we obtain
\begin{align} 
\Tr M^n=(2N)^n \Tr K_{0}+N^n \Tr K_{1} + 2 ^n \Tr K_{2}+ (-2)^n \Tr K_{3}.
\label{trace K2}
\end{align} 
In the large $N$ limit, we obtain by using (\ref{trace K})
\begin{align} 
\Tr M = -N^3, \quad \Tr M^2 = 3N^4, \quad
{\rm Tr}M^n=N^{n+2} \quad (n\ge 3).
\label{trace Mn}
\end{align} 

It is also possible to calculate the effective action for finite $N$ in the $d=0$ model. 
To do this, note that in the loop diagrams including $b_{ab}$, the vector
$b_{ab}$ satisfies the `traceless' condition $ b_{aa}=0$ as in
(\ref{fluct}). This condition changes the propagator for $b_{ab}$
with $M$ replaced by $M'$:
\begin{align} 
M'_{ab,cd}=N K_{1ab,cd} + 2K_{2ab,cd} -2 K_{3ab,cd} .
\label{def M'}
\end{align} 
Hence the $b_{ab}$-loops actually involve $\Tr M^{'n}$ which
are given by
\begin{align} 
\Tr M^{'n}=& N ^n(N^2-1) +  2 ^n \frac{N^2(N+1)(N-3)}{4} + (-2)^n
\frac{N^2(N-1)(N+3)}{4}.
\label{trace Kn}
\end{align} 
Note that it gives the same value to (\ref{trace Mn}) under the
large $N$ limit and the traceless condition indeed affects only finite
$N$ correction.

\section{\label{sec:s-s}Details of the $d=0$ model}

In this Appendix, we will present details pertaining to the $d=0$
model of Section \ref{sec 0d MM}. Specifically, we will
derive \eq{cal-s0-1-val}.

\begin{figure}
\begin{center}
\includegraphics[scale=.9]{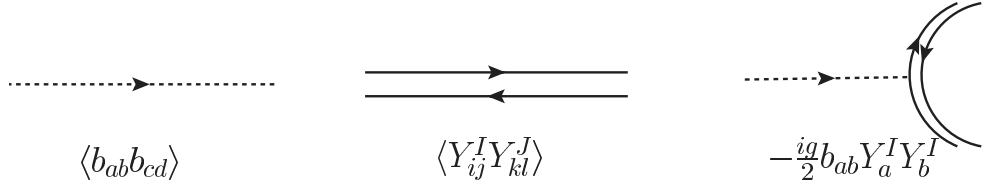}
\caption{Feynman rules for the matrix model \eq{d=0-big}.}
\label{fig tad}
\end{center}
\end{figure}

We start with \eq{d=0-big}.  The Feynman rules of \eq{d=0-big} are
shown in Figure \ref{fig tad}.  The propagators are given by
\begin{align} 
\langle Y^I_a Y^J_b \rangle =\frac{i}{B_0} \delta_{ab}
\delta^{IJ},\qquad \langle b_{ab} b_{cd} \rangle = 2M'_{ab,cd}.
\end{align} 
 The matrix $M'_{ab,cd}$ is defined in (\ref{def M'}), which is the
 propagator for $b_{ab}$ satisfying the traceless condition
 $b_{aa}=0$.  Note that $M'_{ab,cd}$ is obtained by removing $K_0$
 from $M_{ab,cd}$, where $K_0$ is the projection operator
 corresponding to the trace part $b_{aa}$ (See eq.(\ref{projection
   K0})). As remarked below \eq{trace Kn}, the difference between $M$
 and $M'$ appears only at subleading orders in $1/N$.

The effective action ${\cal S}(B_0)$ in \eq{z-b0} is formally given by
\begin{align}
\exp[- {\cal S}(B_0)] &=\int dY^I db
 e^{-S_{0}-S_q} \left(1+\sum_{n=1}^\infty \frac{(-S_{int})^n}{n!}
 \right) \nonumber \\ &=e^{-S_{0}} 
 \left(-iB_0\right)^{-\frac{D(N^2-1)}{2} } \left(1+\sum_{n=1}^\infty
 \frac{ (-)^n }{n!}\langle S_{int}^{n}\rangle \right) .
\label{0d partition function}
\end{align} 
where we have dropped some irrelevant normalization factor.  Thus,
${\cal S}(B_0)$ is given by a sum of all connected vacuum diagrams
represented by the above equation, with external $B_0$.  The term in
the above equation obtained by putting $S_{int}=0$ corresponds to
$S^{(0)}(B_0)$ in \eq{large-D-logic}. Diagrammatically it corresponds
to the $Y^I$-loop in Figure \ref{fig D orders}(a) plus terms
independent of $Y$ and $b$. Since each $Y$-loop gives rise to a factor
$D$, this term is of order $O(D)$, consistent with the arguments in
Section \ref{sec:large-D}.

\begin{figure}
\begin{center}
\includegraphics[scale=0.8]{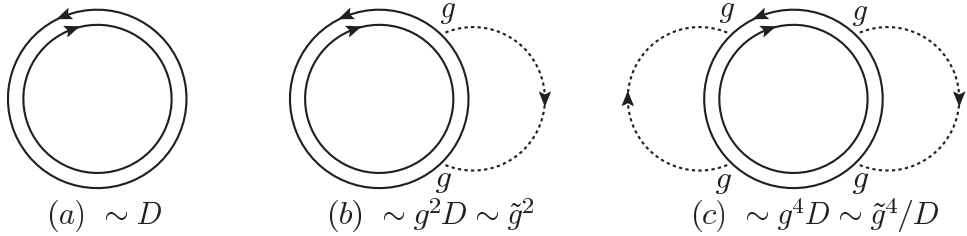}
\caption{Some examples of vacuum diagrams of \eq{0d partition function}.
Figure (a) is the leading order in the $D$ expansion ($O(D)$).
(b) is next order ($O(1)$) and (c) is $O(1/D).$
The diagrams in Figure \ref{fig loop} also contribute to $O(1)$.
 }
\label{fig D orders}
\end{center}
\end{figure}

Therefore, we obtain ${\cal S}_0$ in (\ref{cal-s0-1}) as,
\begin{align} 
{\cal S}_0/D =- \frac{N\triangle^4}{8\tilde{g}^2 }
+\frac{N^2-1}{4}\log \left( \frac{\triangle^4}{\tilde{g}^{2} }
\right) ,
\end{align} 
where $\tilde{g}^2=g^2D$, as defined in (\ref{tilde-g}).  Here we have
used the notation 
\be B_0 = i \triangle^2,
\label{b0-triangle}
\ee in anticipation of the fact that the saddle point value
\eq{condensation 0d} will be given in terms of real $\triangle$. 

\subsection{Calculation of ${\cal S}_1$: the $1/D$ correction}
Now we calculate ${\cal S}_{1}$ in \eq{cal-s0-1}. 

\begin{figure}[h!]
\begin{center}
\includegraphics[scale=.75]{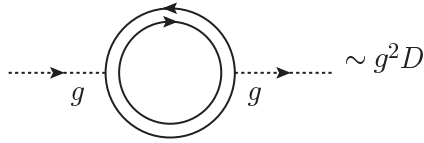}
\caption{One-loop correction to the $\langle bb \rangle$  propagators. 
The correction to $\langle bb \rangle$ is order $\tilde{g}^2$. 
The 1PI propagator comes  from this diagram only at this order in the $1/D$ expansion. }
\label{fig prop}
\end{center}
\end{figure}
We can obtain ${\cal S}_{1}$ directly by using the full propagator for
$\langle bb \rangle$, as in Section \ref{sec:large-D}.  However we
calculate it diagrammatically here\footnote{\label{Dyson} The
  connection with Section \ref{sec:large-D} can be made by the formal
  Schwinger-Dyson sum $S^{(2)}= M + MGM + MGMGM + \cdots $ where $S^{(2)}$
  is the quantity appearing in \eq{large-D-logic}, while $G= G_{(2)}$
  appears in \eq{composite propagator d=0}.}, since this derivation
will be more convenient. It is easy to show that at leading order
in $1/D$, the relevant correction to the 1PI $\vev{bb}$ propagator
comes entirely from the one-loop diagram of Figure \ref{fig
  prop}. Hence only diagrams described in Figure \ref{fig loop}
contribute to the effective action.  Note that each diagram includes
planar and non-planar structures in the $N$ counting.
\begin{figure}
\begin{center}
\includegraphics[scale=.8]{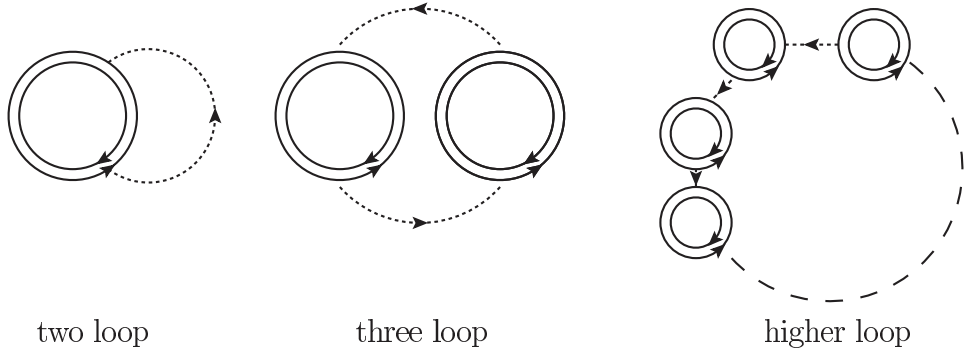}
\caption{The $O(1)$ corrections to the effective action in the
  large $D$ expansion.}
\label{fig loop}
\end{center}
\end{figure}
Higher loop terms are characterized by higher powers of
the dimensionless quantity $\tilde{\lambda}/\triangle^4$.  However, as
evident from \eq{condensation 0d}, this quantity is order 1. Hence we
must sum over all loops, which we describe below.

In each diagram in Figure \ref{fig loop}, the two $Y^I$s in a vertex
$b_{ab} Y^I_a Y^I_b$ are contracted with two $Y^J$s in a different
vertex $b_{cd} Y^J_c Y^J_d$.  Therefore it is convenient to define a
composite propagator\footnote{See footnote \ref{Dyson} 
for another motivation for defining this propagator.}
\begin{align} 
G_{(2)ab,cd}\equiv
\sum_{I,J}
\left( 
\langle Y^I_a Y^J_c  \rangle 
\langle Y^I_b Y^J_d  \rangle +
\langle Y^I_a Y^J_d  \rangle 
\langle Y^I_b Y^J_c  \rangle\right)  = \frac{2D}{\triangle^4}I_{ab,cd} ,
\label{composite propagator d=0}
\end{align} 
where $I_{ab,cd}$ is defined in (\ref{def Iabcd}) and it satisfies
$M'_{ab,cd}I_{cd,ef}=M'_{ab,ef}$.  This composite propagator  corresponds to one double line loop in Figure
\ref{fig prop} and \ref{fig loop}.  By using this propagator, we can
calculate the ($n+1$)-th loop correction to $N^2 {\cal S}_1$ as
\begin{align} 
-\frac{1}{(2n)!}\langle S_{int}^{2n} \rangle_c 
 =&- \frac{1}{(2n)!}\left(\frac{-ig}{2} \right)^{2n} \langle b_{a_1
  b_1} Y_{a_1}^{I_1} Y_{b_1}^{I_2} \cdots b_{a_{2n} b_{2n}}
Y_{a_{2n}}^{I_{2n}} Y_{b_{2n}}^{I_{2n}}\rangle_c \nonumber \\ =& -
\frac{(-)^n}{2n}\left(\frac{g^2D}{\triangle^4} \right)^{n}
M'_{a_1b_1,a_2b_2}I_{a_2b_2, a_3b_3}M'_{a_3b_3,a_4b_4 }\cdots
I_{a_{2n}b_{2n},a_1b_1} \nonumber \\ =& -
\frac{(-)^n}{2n}\left(\frac{g^2D}{\triangle^4} \right)^{n} {\rm Tr}M^{'n}.
\label{s1n}
\end{align} 
Here $\langle \cdots \rangle_c$ denotes the connected diagram. 
$\Tr'M^n$ has been calculated in (\ref{trace Kn}).
Now we can sum over $n$ and obtain the effective action,
\begin{align}
 {\cal S}/D =&- \frac{N\triangle^4}{8\tilde{g}^2 } +\frac{(N^2-1)}{4}\log
 \left( \frac{\triangle^4}{\tilde{g}^2N } \right) + \frac{N^2-1}{2D}
 \log\left(1+\frac{\tilde{g}^2N}{\triangle^4} \right) \nonumber \\ &
 +\frac{N^2(N+1)(N-3)}{8D} \log\left(1+\frac{2\tilde{g}^2}{\triangle^4}
 \right) \nonumber \\ &+\frac{N^2(N-1)(N+3)}{8D}
 \log\left(1-\frac{2\tilde{g}^2}{\triangle^4} \right) +O\left(\frac{1}{D^2}
 \right).
\label{finite N d=0}
\end{align}

\section{Evaluation of a toy integral using a complex saddle point}
\label{saddle-point}

In Section \ref{sec 0d MM}, we evaluated the partition function
\eq{z-b0} by a saddle point method. A similar calculation was done in
Section \ref{sec d=1 contd}.  In this Appendix, we illustrate the
procedure by considering a toy example. 

Let us consider the integral
\begin{align}
I=  \int_{-\infty}^\infty dy e^{-ay^2-by^4} ~~~~~(a,b>0).
\end{align}
This integral can be evaluated by using the Bessel functions.
Alternatively, if $b$ is small we can expand in powers of $b$ 
and obtain,
\begin{align}
I= \sqrt{\frac{\pi}{a}} -\frac{3b\sqrt \pi}{4a^{5/2}} + \cdots .
\end{align}
Now we try to solve this integral by using the auxiliary variable.  We
can rewrite the integral as
\begin{align}
&\frac{1}{\sqrt{\pi b}}   \int_{-\infty}^\infty dy dx
  \exp\left( -ay^2-\frac{x^2}{b}+2ixy^2 \right)  \nonumber \\
  =&\frac{1}{\sqrt{ b}}   \int_{-\infty}^\infty  dx
  \exp\left(-\frac{x^2}{b} -\frac{1}{2}\log(a-2ix)  \right).  
\end{align}
Let us try to evaluate this integral by using the saddle point method,
in the limit $b\to 0$.  The exponent has two saddle points $x=-ia/4
\pm i \sqrt{a^2+4b}/4 $. Since in the limit $b=0$, the extremum is at
$x=0$, we should choose the saddle point corresponding to the ``$+$"
sign.  We get
\begin{align}
I= \sqrt{\frac{\pi}{a}} -\frac{3b\sqrt \pi}{4a^{5/2}} + \cdots ,
\end{align}
reproducing the earlier expression.

\section{The $Y$-Propagator for $d=1$}
\label{app propagator}

In this section, we derive the $Y^I$ propagator in the $d=1$ model.
The kinetic term of $Y^I$ in (\ref{action 1dMM}) can be written as,
\begin{align} 
&\int_0^\beta dt {\rm Tr}\frac{1}{2} \left[Y^I \left( -D_0^2 + \triangle^2 \right)
    Y^I \right] \nonumber \\ &=\frac{\beta}{2} \sum_{i,j,n} Y_{nij}^I
  \left( \frac{4\pi^2 n^2}{\beta^2}-\frac{4\pi n (\alpha_j -
    \alpha_i)}{\beta} + (\alpha_j-\alpha_i)^2+ \triangle^2 \right)
  Y_{-nji}^I,
\end{align} 
where we have used the constant diagonal gauge $A_{0ij}=\alpha_i \delta_{ij}$ and we have expanded
$Y^I(t)=\sum_n Y_n^I e^{2\pi in/\beta} $. We have also used the
notation $B_0 = i \triangle^2$ as in Section \ref{sec d=1
  contd}. Hence the propagator for each mode is given by
\begin{align} 
\langle Y_{nij}^I Y_{m kl}^J \rangle
=\frac{1}{\beta} \frac{1}{\frac{4\pi^2 n^2}{\beta^2}-\frac{4\pi n  
(\alpha_j - \alpha_i)}{\beta} + (\alpha_j-\alpha_i)^2+ \triangle^2 
}  \delta_{il} \delta_{jk} \delta^{IJ} \delta_{n+m,0} .
\end{align} 
Then the propagator for $Y^I(t)$ becomes
\begin{align} 
\langle Y_{ij}^I(t) Y_{kl}^J(0) \rangle
&=\sum_n \frac{1}{\beta} \frac{e^{\frac{ 2\pi i n}{\beta} t }}{
\frac{4\pi^2 n^2}{\beta^2}-\frac{4\pi n  (\alpha_j - \alpha_i)}{\beta} + (
\alpha_j-\alpha_i)^2+ \triangle^2 
}  \delta_{il} \delta_{jk} \delta^{IJ}  \nonumber \\
 &=\sum_n \frac{-i}{4\pi \triangle}
\Biggl[
\frac{e^{\frac{ 2\pi i n}{\beta} t }}{ n- \frac{\beta   (\alpha_j-\alpha_i)}{
2\pi}-i    \frac{\beta \triangle}{2\pi} 
}  
-\frac{e^{\frac{ 2\pi i n}{\beta} t }}{ n- \frac{\beta   (\alpha_j-\alpha_i)}{
2\pi}+i    \frac{\beta \triangle}{2\pi} 
}  
\Biggr]\delta_{il} \delta_{jk} \delta^{IJ} 
 \nonumber \\
   &= \frac{
  e^{ 
  i    (\alpha_j-\alpha_i)||t||
 }
 }{2 \triangle}
\Biggl[
 \frac{e^{
 -  \triangle ||t|| 
  }}{1-e^{i\beta  (\alpha_j-\alpha_i)}e^{-\beta \triangle}}
-\frac{e^{
   \triangle ||t|| 
  }}{1-e^{i\beta  (\alpha_j-\alpha_i)}e^{\beta \triangle}}\Biggr]\delta_{il} 
\delta_{jk} \delta^{IJ} .
\label{propagator with A0}
 \end{align} 
Here  $||t||$ denotes $||t+n \beta||=t$ for $0\le t < \beta$.
In order to derive it, we have used the formulae\footnote{Eqns. (\ref{sum formula}), (\ref{sum formula 2}) and (\ref{sum formula 3}) are shown in \cite{SUGAKU}.},
\begin{align} 
\sum_{n=-\infty}^{\infty}\frac{\sin(a-n)x}{a-n}=\pi,\quad 
\sum_{n=-\infty}^{\infty}\frac{\cos(a-n)x}{a-n}=\pi \cot(\pi a).
\label{sum formula}
\end{align} 

We can write the expression (\ref{propagator with A0}) further as
\begin{align} 
&\langle Y_{ij}^I(t) Y_{kl}^J(0) \rangle= \nonumber \\
&  \frac{1}{2\triangle}\Biggl[
  e^{ 
  (i    (\alpha_j-\alpha_i)-\triangle)||t||
 }
\sum_{n=0}^{\infty}x^n u^j_n u_{-n}^{i}+
  e^{ 
  (-i    (\alpha_j-\alpha_i)-\triangle)(\beta-||t||)
 }
\sum_{n=0}^{\infty}x^n u^{j}_{-n} u_n^{i}
\Biggr]\delta_{il} \delta_{jk} \delta^{IJ} ,
\label{propagator with A}
 \end{align} 
where $x = e^{-\beta \triangle}$ and $u_n^i = e^{i \beta n \alpha_i}$
which satisfies $\sum_{i=1}^{N} u_n^i =N u_n$.

\section{All loop corrections up to $1/D$ in the $d=1$ model}
\label{app all loop}

In this appendix, we will show the derivation of the effective action
of the $d=1$ model including the leading $1/D$ correction
(\ref{s-eff-all}).  This correction corresponds to ${\cal S}_{1,0}$ in
(\ref{double-expansion}) in the 0 dimensional model. Even in $d=1$
model (\ref{action 1dMM}), the same diagrams as in the $0$ dimensional
model (Figure \ref{fig loop}) will contribute.  Therefore, as we
discussed in Appendix \ref{sec:s-s}, the following composite
propagator is convenient to calculate the loops,
\begin{align} 
\sum_{j,p,I,J} \langle Y_{ij}^I(t) Y_{pq}^J(t') \rangle \langle
Y_{jk}^I(t) Y_{lp}^J(t') \rangle \equiv DN \sum_{n}G^{(2)}_{n,ik}
e^{i\frac{2\pi n}{\beta}(t-t') } \delta_{iq}\delta_{kl}.
\label{def 2propagator-app}
\end{align} 
Note that in this propagator we have only taken into account
contractions which corresponds to planar diagrams. It turns out that
the effective action obtained in this way corresponds to the leading
term in a $1/N$ expansion. We will make a brief remark about
non-planar terms at the end of this Appendix.

We can calculate this composite propagator by using (\ref{propagator
  with A}) and a formula for Fourier integrals involving $||t||$,
\begin{align} 
\frac{1}{\beta^2} \int_0^\beta \int_0^\beta dt dt'
e^{s||t-t'||}e^{-\frac{2\pi i n}{\beta}t } 
e^{-\frac{2\pi i m}{\beta}t' }
=\delta_{n+m,0}\frac{e^{s\beta}-1}{s\beta-2\pi i n}.
\end{align} 
Then the composite propagator can be obtained as
\begin{align} 
G^{(2)}_{n,ik}=\frac{1}{8\triangle^2}
\left( P^-_{n,ik}S^-_{ik} + P^+_{n,ik} S^+_{ik}+Q_{n,ik}S_{Q,ik} \right)  ,
\label{sol composite propagator}
\end{align} 
where the $n$-independent quantities are given by
\begin{align} 
S^-_{ik}&=1+\sum_{m=1}^{\infty}x^m( u_{-m}^i u_m+u_m^k u_{-m}), \\
S^+_{ik}&=(S^-_{ik})^*=
1+\sum_{m=1}^{\infty}x^m( u_{m}^i u_{-m}+u_{-m}^k u_{m}),  \\
S_{Q,ik}&=x\sum_{l,m=0}^{\infty}x^{l+m}
[ u_{l+m+1}u_{-l}^i u_{-m}^k(u_{-1}^i-u_{-1}^k)+u_{-(l+m+1)}
u_{l}^i u_{m}^k(u_{1}^k-u_{1}^i)],
\label{def SQ}
\end{align} 
and the $n$-dependent quantities are given by
\begin{align} 
P^-_{n,ik}&=\frac{1}{\pi i}\frac{-1}{\frac{i\beta 
(\alpha_k-\alpha_i)-2\triangle \beta}{2\pi i}-n },\qquad
P^+_{n,ik}=\frac{1}{\pi i}\frac{1}{\frac{i\beta 
(\alpha_k-\alpha_i)+2\triangle \beta}{2\pi i}-n },  \nonumber \\
Q_{n,ik}&=\frac{1}{\pi i}\frac{1}{\frac{i\beta (\alpha_k-\alpha_i)}{
2\pi i}-n } . 
\end{align} 

By using the composite propagator (\ref{def 2propagator-app}), we can
calculate the loop correction to the effective action as we studied in
Appendix \ref{sec:s-s}.  The $(n+1)$-loop correction in Figure
\ref{fig loop} is given by
\begin{align} 
-d_n\frac{(-)^n}{2n}
\left( \beta  g^2D  N\right)^n 
\sum_{m=-\infty}^\infty  \sum_{i,j=1}^N  \left( G^{(2)}_{m,ij} \right)^n, 
\label{effective action general loop}
\end{align} 
where $d_n$ is a factor derived from the number of the planar diagrams
and we can fix it by using (\ref{trace Mn}),
\begin{align} 
d_1=-1,~~d_2=3,~~d_n=1~~(n \ge 3) \nonumber.
\end{align} 
It is difficult to evaluate (\ref{effective action general loop}) in
general.  However, we are interested in the theory around the critical
temperature $\bar{x}\sim 1/D$ where $\bar x$ is given by \eq{bar-x}.
Hence we can expand the effective potential with respect to $x$ and
the lowest order of $x$ is enough to evaluate the $1/D$ correction.
Especially, only the coefficient of $x |u_1|^2$ and $x^2|u_1|^4$ and
the gauge-field independent terms will give us the relevant
information of the dynamics around the critical points.

\subsection{Two-loop correction}

The two-loop correction to the effective action corresponds to the
$n=1$ term in (\ref{effective action general loop}) and is given by
\begin{align} 
-\frac{1}{2}
\beta  g^2D  N
\sum_{i,j=1}^N \sum_{m=-\infty}^\infty    G^{(2)}_{m,ij}.
\end{align} 
We sum over the Fourier mode first. 
\begin{align} 
\sum_{m=-\infty}^\infty P^-_{m,ij}&=1+2\sum_{m=1}^{\infty}x^{2m} 
u_{-m}^i u_{m}^j, \quad
\sum_{m=-\infty}^\infty P^+_{m,ij}=1+2\sum_{m=1}^{\infty}x^{2m} 
u_{m}^i u_{-m}^j ,\nonumber \\
\sum_{m=-\infty}^\infty Q_{m,ij}&=\frac{u_{-1}^i+u_{-1}^j}{u_{-1}^i-u_{-1}^j},
\end{align} 
where we have used (\ref{sum formula}).  After summing over $i,j$, we
obtain the correction to the two-loop effective action as,
\begin{align} 
S^{two-loop}&=-\frac{N^2 \beta\triangle}{8}\left(  
\frac{\tilde{\lambda} }{\triangle^3} \right) 
\left(1+2 \sum_{m=1}^\infty (x^{2m}+2x^m )|u_m|^2  \right) +S_{int}, \\
S_{int}&=-\frac{N^2 \beta\triangle}{4} \left(  \frac{\tilde{\lambda} }{
\triangle^3} \right) 
\left( x^2 (u_1^2 u_{-2}+u_{-1}^2u_2) +O(x^3)\right) .
\label{two-loop interaction}
\end{align} 
Here we have omitted higher $x$ terms in the interaction, which are
irrelevant around the critical points, as argued before.

We notice that this correction includes a cubic interaction $x^2 u_1^2
u_{-2}$.   Since the effective action (\ref{effective action d=1}) has
the term $|u_2|^2/2$ arising from the gauge fixing, $|u_1|^4$ term can
be induced through this interaction after integrating out $u_2$.
However, the coefficient of the $|u_1|^4$ terms obtained
this way will be $O(x^4)$ and we can ignore it compared to the
$|u_1|^4$ potential in (\ref{LG'}).  Generally, we can show that the
lowest order coefficient of the cubic interaction from the higher
loops is also $x^2$ and we will ignore them here.

\subsection{Three-loop correction}
We evaluate the three-loop correction (the $n=2$ term 
in \eq{effective action general loop})
\begin{align} 
-\frac{3}{4}
\left( \beta  g^2D  N\right)^2 
\sum_{m=-\infty}^\infty  \sum_{i,j=1}^N  \left( G^{(2)}_{m,ij} \right)^2.
\end{align} 
In order to calculate the sum of the Fourier mode $m$,
we derive the product of $P^{\pm}$ and $Q$ as
\begin{align} 
P^-_{m,ij}P^+_{m,ij}&=\frac{1}{2  \triangle\beta}\left(P^-_{m,ij}+P^+_{m,ij} 
\right), \nonumber \\
P^-_{m,ij}  Q_{m,ij}&=\frac{1}{ \triangle\beta}\left(P^-_{m,ij}+  Q_{m,ij} 
\right) ,~~
P^+_{m,ij}  Q_{m,ij}=\frac{1}{ \triangle\beta}\left(-P^+_{m,ij}+  Q_{m,ij} 
\right) . 
\label{product PQ}
\end{align} 
We also calculate the sum of squares of these 
quantities by using the formula
\begin{align} 
\sum_{m=-\infty}^{\infty}\left(\frac{1}{a-m}
\right)^2=-\frac{\partial}{ \partial a}
\sum_{m=-\infty}^{\infty}\frac{1}{a-m}=\frac{\pi^2}{\sin^2 \pi a}.
\label{sum trick}
\end{align} 
This leads to 
\begin{align} 
\sum_{m=-\infty}^\infty (P^-_{m,ij})^2&=\frac{4x^2u_{-1}^i u_1^j}{(1-u_{-1}^i 
u_1^jx^2)^2} , \qquad
\sum_{m=-\infty}^\infty \left( P^+_{m,ij}\right)^2=\frac{4x^2u_{1}^i u_{-1}^j}{
(1-u_{1}^i u_{-1}^jx^2)^2}, \nonumber \\
\sum_{m=-\infty}^\infty \left( Q_{m,ij}\right)^2&=\frac{4u_{-1}^i u_1^j}{
(1-u_{-1}^iu_{1}^j)^2} .
\label{P^2}
\end{align} 
Now we can sum over $i,j$ and obtain the leading order of the
corrections as
\begin{align} 
S^{three-loop}=&N^2\Biggl[ -\frac{3}{128}\frac{\beta \tilde{\lambda}^2 }{
\triangle^5}   
+\left( -\frac{9}{32}\frac{\beta \tilde{\lambda}^2}{ 
\triangle^5} x    +O(x^2) \right)  |u_1|^2 
\nonumber \\
 &~~~+\left( -\frac{3}{32}\frac{ \beta \tilde{\lambda}^2 }{\triangle^5} x^2 \left(
\frac{5}{2} +\beta \triangle \right)  +O(x^3)\right)  |u_1|^4 
\Biggr]+\cdots.
\end{align} 

\subsection{$(n+1)$-loop correction to effective potential}

Up to three loops, the leading order of the coefficient of $|u_1|^2$
is $x$ and $|u_1|^4$ is $ x^2$ in the $x$ expansion.  We can find that
these are true even in an arbitrary loop.  Thus it is enough to fix
the coefficient of these terms in each loop.  In order to evaluate the
$(n+1)$-loop, we have to calculate
\begin{align} 
& \sum_{m=-\infty}^\infty  \sum_{i,j=1}^N  \left( G^{(2)}_{m,ij} \right)^n
=&\left( \frac{1}{8\triangle^2}\right) ^2
\sum_{m=-\infty}^\infty  \sum_{i,j=1}^N \left( P^-_{m,ij}S^-_{ij} + 
P^+_{m,ij} S^+_{ij}+Q_{m,ij}S_{Q,ij} \right)^n.
\end{align} 
in \eq{effective action general loop}.  However we can reduce this
equation.  Since $S_{Q,ij}$ is an order $x$ quantity as in (\ref{def
  SQ}), we can ignore $S_Q^k$ ($k \ge 3$) terms here. Then we should
evaluate only
\begin{align} 
\left( P^-S^- + P^+ S^+ \right)^n+n
QS_Q\left( P^-S^- + P^+ S^+ \right)^{n-1} \nonumber \\
+\frac{n(n-1)}{2} \left( QS_{Q} \right)^2 \left( P^-S^- + P^+ S^+\right)^{n-2},
\label{expansion-n-loop}
\end{align} 
where we have omitted the indices.

First we calculate the first term in (\ref{expansion-n-loop}).  It is
convenient to define $a_{k,l}=\left( P^{+}\right)^k \left( P^-
\right)^l$ so that the equation becomes,
\begin{align} 
\left( P^-S^- + P^+ S^+ \right)^n
&=\sum_{k=0}^n \binom nk a_{n-k,k}S_+^{n-k} S_-^{k} .
\label{sum-pp-pm}
\end{align}  
Through the relation (\ref{product PQ}), $a_{k,l}$ satisfy
\begin{align} 
a_{k,l}=\frac{1}{2\triangle \beta}\left(a_{k,l-1}+a_{k-1,l} \right)  .
\label{akl-relation}
\end{align} 
Here we can approximate $a_{k,0}=a_{0,k}=0$ if $k\ge 2$.  This is
because we can show that they are order $x^2$ quantities and the
lowest order terms only contribute to $x^2 |u_1|^2$ by using a similar
logic as in (\ref{sum trick}) and (\ref{P^2}).  Then, through
(\ref{akl-relation}), we can obtain
\begin{align} 
a_{k,l}=&\frac{1}{(2\triangle
  \beta)^{k+l-1}} \sum_{k_{l-1}=1}^{k}\sum_{k_{l-2}=1}^{k_{l-1}} \cdots
  \sum_{k_{2}=1}^{k_{3}}\sum_{k_{1}=1}^{k_{2}}
  \left(a_{1,0}+a_{0,1}
\right) \nonumber \\
=&\frac{1}{(2\triangle
  \beta)^{k+l-1}}\frac{(k+l-2)!}{(k-1)!(l-1)!}\left(a_{1,0}+a_{0,1}
\right) .
\end{align} 
Since $a_{1,0}=P^+_{m,ij}$ and $a_{0,1}=P^-_{m,ij}$, we can sum over
the $i,j$ and $m$ indices in (\ref{sum-pp-pm}) as
\begin{align} 
&\sum_{m=-\infty}^{\infty}\sum_{i,j=1}^N\left(P^+_{m,ij}+P^-_{m,ij}\right) 
\left( S^+_{ij}\right)^{n-k}\left( S^-_{ij}\right)^{k}  \nonumber \\
=&2N^2 \left(1+2n x |u_1|^2+n(n-1)x^2 |u_1|^4 +\cdots \right) ,
\end{align} 
where $\cdots$ denotes the irrelevant higher order terms.
Then (\ref{sum-pp-pm}) becomes
\begin{align} 
&\frac{2^nN^2}{(\triangle \beta)^{n-1}}\frac{(2n-3)!!}{(2n-2)!!}
\left( 1+2nx |u_1|^2  +n(n-1)x^2 |u_1|^4 \right) + \cdots,
\end{align} 
where we have used 
\begin{align} 
\sum_{k=0}^n \binom nk \binom{n-2}{k-1}=2^{2(n-1)}\frac{(2n-3)!!}{(2n-2)!!}.
\label{sum formula 2}
\end{align} 
Next we evaluate 
\begin{align} 
Q S_Q \left( P^-S^- + P^+ S^+ \right)^{n-1},
\end{align} 
in (\ref{expansion-n-loop}).
First we can show that
\begin{align} 
&\sum_{m=-\infty}^{\infty}\sum_{i,j=1}^N\left(P^{\pm}_{m,ij}\right)
  \left( S^+_{ij}\right)^{l}\left( S^-_{ij}\right)^{k} S_{Q,ij} ,
  \nonumber
\end{align} 
does not contribute to the relevant potential.
Hence, by using (\ref{product PQ}), we obtain,
\begin{align} 
&\sum_{m=-\infty}^{\infty}\sum_{i,j=1}^N Q_{m,ij} S_{Q,ij} \left( 
P_{m,ij}^-S^-_{ij} + P_{m,ij}^+ S^+_{ij} \right)^{n-1} \nonumber \\
=&\frac{1}{(\triangle \beta)^{n-1}} \sum_{m=-\infty}^{\infty}
\sum_{i,j=1}^N Q_{m,ij} S_{Q,ij} \left( S^-_{ij} + S^+_{ij} \right)^{n-1}+
\cdots \nonumber \\
=&\frac{4 N^2 2^{n-1}}{(\triangle \beta)^{n-1}} 
\left(x|u_1|^2+(n-1)x^2 |u_1|^4 \right) +\cdots.
\end{align} 
Finally we evaluate
\begin{align} 
\left( Q S_Q \right)^2 \left( P^-S^- + P^+ S^+ \right)^{n-2}.
\end{align} 
Here we can show that
\begin{align} 
&\sum_{m=-\infty}^{\infty}\sum_{i,j=1}^N\left(P^{\pm}_{m,ij}\right) 
\left( S^+_{ij}\right)^k\left( S^-_{ij}\right)^{l} S_{Q,ij}^2 , \nonumber
\end{align} 
does not contribute to the relevant potential.
Thus we obtain
\begin{align} 
&\sum_{m=-\infty}^{\infty}\sum_{i,j=1}^N \left(  Q_{m,ij} S_{Q,ij}\right)^2  
\left( P_{m,ij}^-S^-_{ij} + P_{m,ij}^+ S^+_{ij} \right)^{n-2} \nonumber \\
=&\frac{2^{n-2}}{(\triangle \beta)^{n-2}} \sum_{m=-\infty}^{\infty}
\sum_{i,j=1}^N \left(  Q_{m,ij} S_{Q,ij}\right)^2 +\cdots 
=\frac{2^{n-2}}{(\triangle \beta)^{n-2}} 
\left(8N^2 x^2 |u_1|^4 \right)  +\cdots.
\end{align} 

Let us summarize all relevant terms of the $(n+1)$-loop effective
action.  The gauge-field independent term becomes
\begin{align} 
-(-)^n d_n N^2 \beta \triangle 
\left( \frac{\tilde{\lambda} }{4 \triangle^3} \right)^n
\frac{(2n-3)!!}{(2n)!!} .
\label{n+1 constant}
\end{align} 
Note that this result is exact.  The leading $|u_1|^2$ potential in
the $x$ expansion is given by
\begin{align} 
-(-)^n d_n N^2 \beta \triangle 
\left( \frac{\tilde{\lambda} }{4 \triangle^3} \right)^n
\left( 
\frac{(2n-3)!!}{(2n-2)!!} 
+1 \right) x|u_1|^2+O(x^2).
\label{n+1 u2}
\end{align} 
The leading $|u_1|^4$ potential is 
\begin{align} 
-(-)^n d_n N^2  \beta \triangle 
\frac{n-1}{2} 
\left( \frac{\tilde{\lambda} }{4 \triangle^3} \right)^n
\left( 
\frac{(2n-3)!!}{(2n-2)!!}+2+\triangle \beta 
 \right) x^2|u_1|^4+O(x^3).
 \label{n+1 u4}
\end{align} 

We can sum over $n$ by using the following formula:
\begin{align}
\sum_{n=1}^\infty (-)^{n-1} \frac{(2n-3)!!}{(2n)!!}x^n=\sqrt{1+x}-1  ,
\label{sum formula 3}
\end{align} 
and its derivative with respect to $x$. With this, we finally obtain
the effective action (\ref{s-eff-all}).

Note that it is possible to extend the calculation in this section to
finite $N$ case as we have done in the $d=0$ model.  The finite $N$
result for the gauge-field independent constant term is simply
obtained by replacing $d_nN^{n+2}$ in (\ref{n+1 constant}) with $\Tr
M^{'n}$ in (\ref{trace Kn}).  The terms including the gauge potential
are more complicated and we have to modify the composite propagator
(\ref{def 2propagator-app}).


\begin{thebibliography}{999}

%\cite{Brezin:1994eb}
\bibitem{Brezin:1994eb}
  E.~Brezin and S.~R.~Wadia,
  ``The Large N expansion in quantum field theory and statistical physics: From
  spin systems to two-dimensional gravity,''
%\href{http://www.slac.stanford.edu/spires/find/hep/www?irn=3134423}{SPIRES entry}
{\it  Singapore, Singapore: World Scientific (1993) 1130 p}

%\cite{Eguchi:1982nm}
\bibitem{Eguchi:1982nm}
  T.~Eguchi and H.~Kawai,
  ``Reduction Of Dynamical Degrees Of Freedom In The Large N Gauge Theory,''
  Phys.\ Rev.\ Lett.\  {\bf 48}, 1063 (1982).
  %%CITATION = PRLTA,48,1063;%%

%\cite{Krauth:1999qw}
\bibitem{Krauth:1999qw} W.~Krauth and M.~Staudacher,
  ``Eigenvalue distributions in Yang-Mills integrals,''
  Phys.\ Lett.\  B {\bf 453}, 253 (1999)
  [arXiv:hep-th/9902113].
  %%CITATION = PHLTA,B453,253;%%

%\cite{Banks:1996vh}
\bibitem{Banks:1996vh}
  T.~Banks, W.~Fischler, S.~H.~Shenker and L.~Susskind,
  ``M theory as a matrix model: A conjecture,''
  Phys.\ Rev.\  D {\bf 55}, 5112 (1997)
  [arXiv:hep-th/9610043].
  %%CITATION = PHRVA,D55,5112;%%

%\cite{Ishibashi:1996xs}
\bibitem{Ishibashi:1996xs}
  N.~Ishibashi, H.~Kawai, Y.~Kitazawa and A.~Tsuchiya,
  ``A large-N reduced model as superstring,''
  Nucl.\ Phys.\  B {\bf 498}, 467 (1997)
  [arXiv:hep-th/9612115].
  %%CITATION = NUPHA,B498,467;%%

%\cite{Hotta:1998en}
\bibitem{Hotta:1998en}
  T.~Hotta, J.~Nishimura and A.~Tsuchiya,
  ``Dynamical aspects of large N reduced models,''
  Nucl.\ Phys.\  B {\bf 545}, 543 (1999)
  [arXiv:hep-th/9811220].
  %%CITATION = NUPHA,B545,543;%%

%\cite{Itzhaki:1998dd}
\bibitem{Itzhaki:1998dd}
  N.~Itzhaki, J.~M.~Maldacena, J.~Sonnenschein and S.~Yankielowicz,
   ``Supergravity and the large N limit of theories with sixteen
  supercharges,''
  Phys.\ Rev.\  D {\bf 58}, 046004 (1998)
  [arXiv:hep-th/9802042].
  %%CITATION = PHRVA,D58,046004;%%


%\cite{Kabat:2000zv}
\bibitem{Kabat:2000zv}
  D.~N.~Kabat, G.~Lifschytz and D.~A.~Lowe,
  ``Black hole thermodynamics from calculations in strongly coupled gauge
  theory,''
  Int.\ J.\ Mod.\ Phys.\  A {\bf 16}, 856 (2001)
  [Phys.\ Rev.\ Lett.\  {\bf 86}, 1426 (2001)]
  [arXiv:hep-th/0007051].
  %%CITATION = PRLTA,86,1426;%%

%\cite{Kabat:2001ve}
\bibitem{Kabat:2001ve}
  D.~N.~Kabat, G.~Lifschytz and D.~A.~Lowe,
  ``Black hole entropy from non-perturbative gauge theory,''
  Phys.\ Rev.\  D {\bf 64}, 124015 (2001)
  [arXiv:hep-th/0105171].
  %%CITATION = PHRVA,D64,124015;%%

%\cite{Anagnostopoulos:2007fw}
\bibitem{Anagnostopoulos:2007fw}
  K.~N.~Anagnostopoulos, M.~Hanada, J.~Nishimura and S.~Takeuchi,
  ``Monte Carlo studies of supersymmetric matrix quantum mechanics with sixteen
  supercharges at finite temperature,''
  Phys.\ Rev.\ Lett.\  {\bf 100}, 021601 (2008)
  [arXiv:0707.4454 [hep-th]].
  %%CITATION = PRLTA,100,021601;%%

%\cite{Hanada:2008gy}
\bibitem{Hanada:2008gy}
  M.~Hanada, A.~Miwa, J.~Nishimura and S.~Takeuchi,
  ``Schwarzschild radius from Monte Carlo calculation of the Wilson loop in
  supersymmetric matrix quantum mechanics,''
  Phys.\ Rev.\ Lett.\  {\bf 102} (2009) 181602
  [arXiv:0811.2081 [hep-th]].
  %%CITATION = PRLTA,102,181602;%%


%\cite{Hanada:2008ez}
\bibitem{Hanada:2008ez}
  M.~Hanada, Y.~Hyakutake, J.~Nishimura and S.~Takeuchi,
  ``Higher derivative corrections to black hole thermodynamics from
  supersymmetric matrix quantum mechanics,''
  Phys.\ Rev.\ Lett.\  {\bf 102} (2009) 191602
  [arXiv:0811.3102 [hep-th]].
  %%CITATION = PRLTA,102,191602;%%


%\cite{Catterall:2008yz}
\bibitem{Catterall:2008yz}
  S.~Catterall and T.~Wiseman,
  ``Black hole thermodynamics from simulations of lattice Yang-Mills theory,''
  Phys.\ Rev.\  D {\bf 78}, 041502 (2008)
  [arXiv:0803.4273 [hep-th]].
  %%CITATION = PHRVA,D78,041502;%%

%\cite{Azeyanagi:2008mi}
\bibitem{Azeyanagi:2008mi}
  T.~Azeyanagi, M.~Hanada, H.~Kawai and Y.~Matsuo,
  ``Worldsheet Analysis of Gauge/Gravity Dualities,''
  Nucl.\ Phys.\  B {\bf 816}, 278 (2009)
  [arXiv:0812.1453 [hep-th]].
  %%CITATION = NUPHA,B816,278;%%

%\cite{Sundborg:1999ue}
\bibitem{Sundborg:1999ue}
  B.~Sundborg,
  ``The Hagedorn Transition, Deconfinement and N=4 SYM Theory,''
  Nucl.\ Phys.\  B {\bf 573}, 349 (2000)
  [arXiv:hep-th/9908001].
  %%CITATION = NUPHA,B573,349;%%


%\cite{Aharony:2003sx}
\bibitem{Aharony:2003sx} 
  O.~Aharony, J.~Marsano, S.~Minwalla,
  K.~Papadodimas and M.~Van Raamsdonk, ``The Hagedorn / deconfinement
  phase transition in weakly coupled large N gauge theories,''
  Adv.\ Theor.\ Math.\ Phys.\ {\bf 8}, 603 (2004)
  [arXiv:hep-th/0310285].
  %%CITATION = 00203,8,603;%%


%\cite{AlvarezGaume:2005fv}
\bibitem{AlvarezGaume:2005fv}
  L.~Alvarez-Gaume, C.~Gomez, H.~Liu and S.~Wadia,
  ``Finite temperature effective action, AdS(5) black holes, and 1/N
  expansion,''
  Phys.\ Rev.\  D {\bf 71}, 124023 (2005)
  [arXiv:hep-th/0502227].
  %%CITATION = PHRVA,D71,124023;%%

%\cite{AlvarezGaume:2006jg}
\bibitem{AlvarezGaume:2006jg}
  L.~Alvarez-Gaume, P.~Basu, M.~Marino and S.~R.~Wadia,
  ``Blackhole / string transition for the small Schwarzschild blackhole of
  AdS(5) x S**5 and critical unitary matrix models,''
  Eur.\ Phys.\ J.\  C {\bf 48}, 647 (2006)
  [arXiv:hep-th/0605041].
  %%CITATION = EPHJA,C48,647;%%

%\cite{Berenstein:2002jq}
\bibitem{Berenstein:2002jq}
  D.~E.~Berenstein, J.~M.~Maldacena and H.~S.~Nastase,
  ``Strings in flat space and pp waves from N = 4 super Yang Mills,''
  JHEP {\bf 0204}, 013 (2002)
  [arXiv:hep-th/0202021].
  %%CITATION = JHEPA,0204,013;%%


%\cite{Klebanov:2003km}
\bibitem{Klebanov:2003km}
  I.~R.~Klebanov, J.~M.~Maldacena and N.~Seiberg,
  ``D-brane decay in two-dimensional string theory,''
  JHEP {\bf 0307}, 045 (2003)
  [arXiv:hep-th/0305159].
  %%CITATION = JHEPA,0307,045;%%

%\cite{McGreevy:2003kb}
\bibitem{McGreevy:2003kb}
  J.~McGreevy and H.~L.~Verlinde,
  ``Strings from tachyons: The c = 1 matrix reloaded,''
  JHEP {\bf 0312}, 054 (2003)
  [arXiv:hep-th/0304224].
  %%CITATION = JHEPA,0312,054;%%

%\cite{Aharony:2004ig}
\bibitem{Aharony:2004ig}
  O.~Aharony, J.~Marsano, S.~Minwalla and T.~Wiseman,
  ``Black hole-black string phase transitions in thermal 1+1-dimensional
  supersymmetric Yang-Mills theory on a circle,''
  Class.\ Quant.\ Grav.\  {\bf 21}, 5169 (2004)
  [arXiv:hep-th/0406210].
  %%CITATION = CQGRD,21,5169;%%

%\cite{Aharony:2005ew}
\bibitem{Aharony:2005ew}
  O.~Aharony, J.~Marsano, S.~Minwalla, K.~Papadodimas, M.~Van Raamsdonk 
  and T.~Wiseman,
  ``The phase structure of low dimensional large N gauge theories on tori,''
  JHEP {\bf 0601}, 140 (2006)
  [arXiv:hep-th/0508077].
  %%CITATION = JHEPA,0601,140;%%


%\cite{Kawahara:2007fn}
\bibitem{Kawahara:2007fn}
  N.~Kawahara, J.~Nishimura and S.~Takeuchi,
  ``Phase structure of matrix quantum mechanics at finite temperature,''
  JHEP {\bf 0710}, 097 (2007)
  [arXiv:0706.3517 [hep-th]].
  %%CITATION = JHEPA,0710,097;%%

%\cite{Azeyanagi:2009zf}
\bibitem{Azeyanagi:2009zf}
  T.~Azeyanagi, M.~Hanada, T.~Hirata and H.~Shimada,
  %``On the shape of a D-brane bound state and its topology change,''
  JHEP {\bf 0903} (2009) 121
  [arXiv:0901.4073 [hep-th]].
  %%CITATION = JHEPA,0903,121;%%

%\cite{Gross:1980he}
\bibitem{Gross:1980he}
  D.~J.~Gross and E.~Witten,
  ``Possible Third Order Phase Transition In The Large N Lattice Gauge
  Theory,''
  Phys.\ Rev.\  D {\bf 21}, 446 (1980).
  %%CITATION = PHRVA,D21,446;%%


%\cite{Wadia:1980cp}
\bibitem{Wadia:1980cp}
  S.~R.~Wadia,
  ``N = Infinity Phase Transition In A Class Of Exactly Soluble Model Lattice
  Gauge Theories,''
  Phys.\ Lett.\  B {\bf 93}, 403 (1980).
  %%CITATION = PHLTA,B93,403;%%

%\cite{Wadia:1979vk Wadia:1979vk}
\bibitem{Wadia:1979vk} S.~Wadia, 
  ``A Study Of U(N) Lattice Gauge Theory In Two-Dimensions,'' 
  preprint EFI-79/44-CHICAGO.
  %%CITATION = EFI-79/44-CHICAGO;%%

%\cite{Drouffe:1983fv}
\bibitem{Drouffe:1983fv}
  J.~M.~Drouffe and J.~B.~Zuber,
  %``Strong Coupling And Mean Field Methods In Lattice Gauge Theories,''
  Phys.\ Rept.\  {\bf 102} (1983) 1.
  %%CITATION = PRPLC,102,1;%%

\bibitem{progress} G.~Mandal, M.~Mahato, T.~Morita and S.~R.~Wadia,
in progress.


%\cite{Jevicki:1979mb}
\bibitem{Jevicki:1979mb}
  A.~Jevicki and B.~Sakita,
  ``The quantum collective field method and its application to the planar
  limit,''
  Nucl.\ Phys.\  B {\bf 165}, 511 (1980).
  %%CITATION = NUPHA,B165,511;%%

%\cite{Jurkiewicz:1982iz}
\bibitem{Jurkiewicz:1982iz}
  J.~Jurkiewicz and K.~Zalewski,
  ``Vacuum Structure Of The U(N $\to$ Infinity) Gauge Theory On A
  Two-Dimensional Lattice For A Broad Class Of Variant Actions,''
  Nucl.\ Phys.\  B {\bf 220}, 167 (1983).
  %%CITATION = NUPHA,B220,167;%%



%\cite{Mandal:1989ry}
\bibitem{Mandal:1989ry}
  G.~Mandal,
  ``Phase structure of unitary matrix models,''
  Mod.\ Phys.\ Lett.\  A {\bf 5}, 1147 (1990).
  %%CITATION = MPLAE,A5,1147;%%

%\cite{Gregory:1993vy}
\bibitem{Gregory:1993vy}
  R.~Gregory and R.~Laflamme,
  ``Black strings and p-branes are unstable,''
  Phys.\ Rev.\ Lett.\  {\bf 70}, 2837 (1993)
  [arXiv:hep-th/9301052].
  %%CITATION = PRLTA,70,2837;%%

%\cite{Kol:2004ww}
\bibitem{Kol:2004ww}
  B.~Kol,
  ``The Phase Transition between Caged Black Holes and Black Strings - A
  Review,''
  Phys.\ Rept.\  {\bf 422}, 119 (2006)
  [arXiv:hep-th/0411240].
  %%CITATION = PRPLC,422,119;%%

%\cite{Taylor:1996ik}
\bibitem{Taylor:1996ik}
  W.~Taylor,
  ``D-brane field theory on compact spaces,''
  Phys.\ Lett.\  B {\bf 394}, 283 (1997)
  [arXiv:hep-th/9611042].
  %%CITATION = PHLTA,B394,283;%%



%\cite{Susskind:1997dr}
\bibitem{Susskind:1997dr}
  L.~Susskind,
  ``Matrix theory black holes and the Gross Witten transition,''
  arXiv:hep-th/9805115.
  %%CITATION = HEP-TH/9805115;%%




%\cite{Slansky:1981yr}
\bibitem{Slansky:1981yr}
  R.~Slansky,
  ``Group Theory For Unified Model Building,''
  Phys.\ Rept.\  {\bf 79}, 1 (1981).
  %%CITATION = PRPLC,79,1;%%


\bibitem{SUGAKU}
S. ~Moriguchi, K.~Udagawa and S.~Hitotsumatsu,
 ``Sugaku Koushiki II,'' 
{\it Iwanami, Tokyo (in Japanese), (2001) 340 p}


\end{thebibliography}
\end{document}